\documentclass[aps,prx,twocolumn,noshowpacs,amsfonts,amssymb,amsmath,longbibliography,superscriptaddress,notitlepage]{revtex4-1}
\usepackage{mathtools}
\usepackage{amssymb}
\usepackage{bbold}
\usepackage{graphicx}
\usepackage[colorlinks=true,linkcolor=blue,urlcolor=blue,citecolor=blue,anchorcolor=blue]{hyperref}
\usepackage{multirow}
\usepackage{color}
\usepackage{booktabs}
\usepackage{algorithm}
\usepackage{algpseudocode}
\usepackage[shortlabels]{enumitem}
\usepackage{relsize}

\usepackage{pdfpages}
\makeatletter
\AtBeginDocument{\let\LS@rot\@undefined}
\makeatother

\AtBeginDocument{
\heavyrulewidth=.08em
\lightrulewidth=.05em
\cmidrulewidth=.03em
\belowrulesep=.65ex
\belowbottomsep=0pt
\aboverulesep=.4ex
\abovetopsep=0pt
\cmidrulesep=\doublerulesep
\cmidrulekern=.5em
\defaultaddspace=.5em
}

\newcommand{\smallstar}{\mathsmaller{\bigstar}}

\newcommand{\pr}[1]{\left(#1 \right)} 
\newcommand{\br}[1]{\left[#1 \right]} 
\newcommand{\cbrace}[1]{\left\{#1 \right\}} 

\newcommand{\avg}[1]{\left< #1 \right>} 
\renewcommand{\d}[2]{\frac{\mathrm{d} #1}{\mathrm{d} #2}} 
\newcommand{\dd}[2]{\frac{\mathrm{d}^2 #1}{\mathrm{d} #2^2}} 
\newcommand{\pd}[2]{\frac{\partial #1}{\partial #2}} 

\let\basetop=\top
\renewcommand{\top}{{}^\basetop \!}

\begin{document}

\title{Influential groups for seeding and sustaining nonlinear contagion\\ in heterogeneous hypergraphs}
    
\author{Guillaume St-Onge}
\email[]{guillaume.st-onge.4@ulaval.ca}
\affiliation{D\'{e}partement de physique, de g\'{e}nie physique et d'optique, Universit\'{e} Laval, Qu\'ebec (Qu\'ebec), Canada, G1V 0A6}
\affiliation{Centre interdisciplinaire en mod\'{e}lisation math\'{e}matique, Universit\'{e} Laval, Qu\'ebec (Qu\'ebec), Canada, G1V 0A6}

\author{Iacopo Iacopini}
\affiliation{Department of Network and Data Science, Central European University, 1100 Vienna, Austria}
\affiliation{Aix Marseille Univ, Universit\'e de Toulon, CNRS, CPT, Marseille, France}
\affiliation{Centre for Advanced Spatial Analysis, University College London, London, W1T 4TJ, United Kingdom}
\affiliation{School of Mathematical Sciences, Queen Mary University of London. Mile End Road, London E1 4NS, United Kingdom}

\author{Vito Latora}
\affiliation{School of Mathematical Sciences, Queen Mary University of London. Mile End Road, London E1 4NS, United Kingdom}
\affiliation{Dipartimento di Fisica ed Astronomia, Università di Catania and INFN, I-95123 Catania, Italy}
\affiliation{Complexity Science Hub, Josefst\"{a}dter Strasse 39, A 1080 Vienna, Austria}

\author{Alain Barrat}
\affiliation{Aix Marseille Univ, Universit\'e de Toulon, CNRS, CPT, Marseille, France}
\affiliation{Tokyo Tech World Research Hub Initiative (WRHI), Tokyo Institute of Technology, Tokyo, Japan}

\author{Giovanni Petri}
\affiliation{Mathematics and Complex Systems Research Area, ISI Foundation, Turin, Italy}
\affiliation{ISI Global Science Foundation, New York, USA}

\author{Antoine Allard}
\affiliation{D\'{e}partement de physique, de g\'{e}nie physique et d'optique, Universit\'{e} Laval, Qu\'ebec (Qu\'ebec), Canada, G1V 0A6}
\affiliation{Centre interdisciplinaire en mod\'{e}lisation math\'{e}matique, Universit\'{e} Laval, Qu\'ebec (Qu\'ebec), Canada, G1V 0A6}
\affiliation{Vermont Complex Systems Center, University of Vermont, Burlington, VT 05401, USA}

\author{Laurent H\'ebert-Dufresne}
\email[]{Laurent.Hebert-Dufresne@uvm.edu}
\affiliation{D\'{e}partement de physique, de g\'{e}nie physique et d'optique, Universit\'{e} Laval, Qu\'ebec (Qu\'ebec), Canada, G1V 0A6}
\affiliation{Vermont Complex Systems Center, University of Vermont, Burlington, VT 05401, USA}
\affiliation{Department of Computer Science, University of Vermont, Burlington, VT 05401, USA}

\date{\today}

\begin{abstract}
    Several biological and social contagion phenomena, such as superspreading events or social reinforcement, are the results of multi-body interactions, for which hypergraphs offer a natural mathematical description.
    In this paper, we develop a novel mathematical framework based on approximate master equations to study contagions on random hypergraphs with a heterogeneous structure, both in terms of group size (hyperedge cardinality) and of membership of nodes to groups (hyperdegree).
    The characterization of the inner dynamics of groups provides an accurate description of the contagion process, without losing the analytical tractability.
    Using a contagion model where multi-body interactions are mapped onto a nonlinear infection rate, our two main results show how large groups are influential, in the sense that they drive both the early spread of a contagion and its endemic state (i.e., its stationary state).
    First, we provide a detailed characterization of the phase transition, which can be continuous or discontinuous with a bistable regime, and derive analytical expressions for the critical and tricritical points.
    We find that large values of the third moment of the membership distribution suppress the emergence of a discontinuous phase transition.
    Furthermore, the combination of heterogeneous group sizes and nonlinear contagion facilitates the onset of a mesoscopic localization phase, where contagion is sustained only by the largest groups, thereby inhibiting bistability as well.
    Second, we formulate the problem of optimal seeding for hypergraph contagion, and we compare two strategies: tuning the allocation of seeds according to either individual node  or group properties. We find that, when the contagion is sufficiently nonlinear, groups are more effective seeds of contagion than individual nodes.
\end{abstract}

\maketitle


\section{Introduction}

Mathematical models of contagion processes enhance our understanding of spreading dynamics and our predictive capabilities \cite{pastor-satorras2015epidemic}.
To account for the interconnected nature of real-world systems, the last two decades of network science and computational epidemiology research have focused on modeling frameworks of increasing complexity~\cite{barrat2008dynamical,pastor-satorras2015epidemic,kiss2017mathematics}.
From the spreading of infectious diseases to rumors and innovations \cite{daley1964epidemics, moreno2004dynamics, rogers2010diffusion}, a crucial aspect remains the interplay between the contact patterns enabling transmission and the dynamics that unfolds on top.
As a representation for these contacts, graphs have been widely exploited to better represent real-world patterns with increasing levels of accuracy~\cite{barrat2008dynamical,pastor-satorras2015epidemic,newman2018networks}.

While graphs remain a reference for the representation of complex systems, they come with a fundamental limitation: they can only encode pairwise interactions (represented by edges).
Groups are instead the foundational units of many biological, ecological and social systems, whose processes can involve multi-body interactions between any number of elements.
To overcome this limitation, higher-order network representations~\cite{torres2020why} are becoming more and more popular~\cite{salnikov2018simplicial, lambiotte2019networks, battiston2020networks} to describe the structure of interacting systems \cite{petri2018simplicial,cencetti2021temporal}, their growth \cite{hebert-dufresne2011structural, young2016growing,bianconi2017emergent,courtney2017weighted} and dynamics in groups \cite{house2008deterministic, hebert-dufresne2010propagation, osullivan2015mathematical}.
Simplicial complexes and hypergraphs can encode relationships and interactions between any number of elements.
They have been used to investigate the implications of higher-order interactions for landmark dynamical processes like synchronization~\cite{bick2016chaos, skardal2019abrupt, millan2020explosive}, diffusion~\cite{schaub2020random, carletti2020random, torres2020simplicial}, and other social dynamics \cite{neuhauser2020multibody, alvarez-rodriguez2021evolutionary, iacopini2021vanishing}.
Ref.~\cite{battiston2020networks} provides a comprehensive review of early efforts in this direction.

Recently, a model of simplicial contagion has been proposed~\cite{iacopini2019simplicial}.
This is a standard Susceptible-Infected-Susceptible (SIS) compartmental model in which a susceptible individual can become infected through different transmission channels, beyond infectious edges.
In models of {\it simple contagion}, the transition from susceptible to infected happens independently with each exposure to an infectious edge \cite{pastor-satorras2015epidemic}.
In models of {\it complex contagion} instead~\cite{lehmann2018complex}, the transition requires multiple infectious edges or is reinforced by multiple exposures, thus accounting for the empirically observed mechanisms of social influence \cite{centola2010spread, karsai2014complex, hodas2014simple, monsted2017evidence}.
In simplicial contagions, or more generally {\it hypergraph contagions}, a susceptible individual can become infected because of a multi-body process, i.e., through an exposure to an infectious group~\cite{iacopini2019simplicial}.
In this way, the node is simultaneously exposed to the state of the entire group, whose effect can be interpreted as a mechanism of social reinforcement \cite{centola2010spread}.
In addition, the study of higher-order contagion models has applications as well in biological sciences:
it has indeed recently been shown that the combination of multi-body interactions, heterogeneous temporal activity, and the concept of a minimal infective dose lead to nonlinear infection kernels in a model of biological contagions~\cite{st-onge2021bursty}.

The analytical approaches derived so far have confirmed the rich phenomenology emerging from the contagion dynamics, characterized by discontinuous phase transitions, bistability and critical mass effects~\cite{iacopini2019simplicial, barrat2021social, jhun2019simplicial, landry2020effect, ferrazdearruda2021phase, cisneros-velarde2020multigroup}.
Most approaches follow a heterogeneous mean-field (HMF) framework in which nodes are divided into hyperdegree classes. These descriptions are analytically tractable, but do not consider the details of the structure and ignore the \textit{dynamical correlations} within groups, which are especially important for hypergraph contagions since multi-body interactions naturally reinforce these correlations.

Other approaches like quenched mean-field theory (QMF) \cite{ferrazdearruda2020social} and microscopic Markov-chains (MMCA)~\cite{matamalas2020abrupt} can explicitly take the entire contact patterns into account.
Along this line, the microscopic epidemic clique equations (MECLE) capture dynamical correlations as well, thereby highlighting the important impact these correlations have on critical points~\cite{burgio2021network}.
The analytical tractability of these approaches is, however, sacrificed in favor of a more precise description of the structure.
To fully understand the consequences of multi-body interactions in contagions on higher-order networks, we thus need a framework that is both analytically tractable and captures dynamical correlations.

In particular, we are interested in better understanding the notion of influence in hypergraph contagions. In classic contagion models on random networks, individual hubs are influential in the sense that they are the best seeds of contagions, but they are also the most apt at sustaining seeded contagions \cite{pastor-satorras2018eigenvector}. However, in hypergraph contagions we must consider the influence of both individuals and groups, because dynamical correlations can allow groups to be more influential than sets of uncorrelated hubs. Regimes of bistability and hysteresis also imply a potential decoupling between the ability of nodes to seed a contagion and their ability to sustain it. We thus wish to determine (i) which set of groups can best maintain the stationary state of a hypergraph contagion, and when this becomes a dominant effect (ii) which set of groups and their configuration offer optimal initial conditions for a contagion, as compared to the classic notion of influential spreaders as individual hubs, and (iii) whether or not these two notions of group influence are aligned.

We use approximate master equations (AMEs)~\cite{hebert-dufresne2010propagation,marceau2010adaptive,gleeson2011highaccuracy,osullivan2015mathematical,st-onge2021social,st-onge2021master} to study hypergraph contagions, capturing exactly the inner dynamics of groups.
We apply this framework to a contagion model where the infection rate is a nonlinear function of the number of infected nodes in groups, and show that \textit{influential groups} can drive both the stationary state of the contagion and its behavior in the transient state.

The paper is structured as follows. In Sec.~\ref{sec:group-based}, we map the simplicial contagion model onto a hypergraph contagion process and introduce the mathematical formalism of group-based AMEs~\cite{hebert-dufresne2010propagation,st-onge2021social,st-onge2021master}.
To illustrate the accuracy of our framework and assess its limitations, we compare the predictions of our AMEs with simulations on real-world hypergraphs and their randomized counterparts.

In Sec.~\ref{sec:phase-transition}, we characterize the importance of influential groups on the phase transition.
We derive analytical expressions for the critical and tricritical points, and study the impact of a heterogeneous structure.
We show that large values for the third moment of the \textit{membership} (hyperdegree) distribution suppress the emergence of a discontinuous phase transition, a result consistent with HMF theories~\cite{jhun2019simplicial,landry2020effect}.
Our formalism allows us to go further and capture the emergence of a \textit{mesoscopic localization} phase, where infected nodes are concentrated in the largest groups \cite{st-onge2021social,st-onge2021master}.
We show that localization effects can drive the onset of an endemic phase and, incidentally, inhibit bistability.

Finally, in Sec.~\ref{sec:influence_maximization}, we showcase the importance of influential groups on the transient state by considering a problem of influence maximization.
We define and solve the optimization problem based on two strategies, allocating seeds according to either node individual properties or according to group properties.
When the contagion is sufficiently nonlinear, we find that the latter is more effective. This suggests again that contagion processes driven by \textit{groups} lead to a fundamentally different phenomenology than
pairwise contagions, such that investigations and modeling of social contagions should carefully account for higher-order interactions.



\section{Results}

\subsection{Hypergraph contagion model}
\label{sec:group-based}

\begin{figure}[tb]
\begin{center}
    \includegraphics[width=\linewidth]{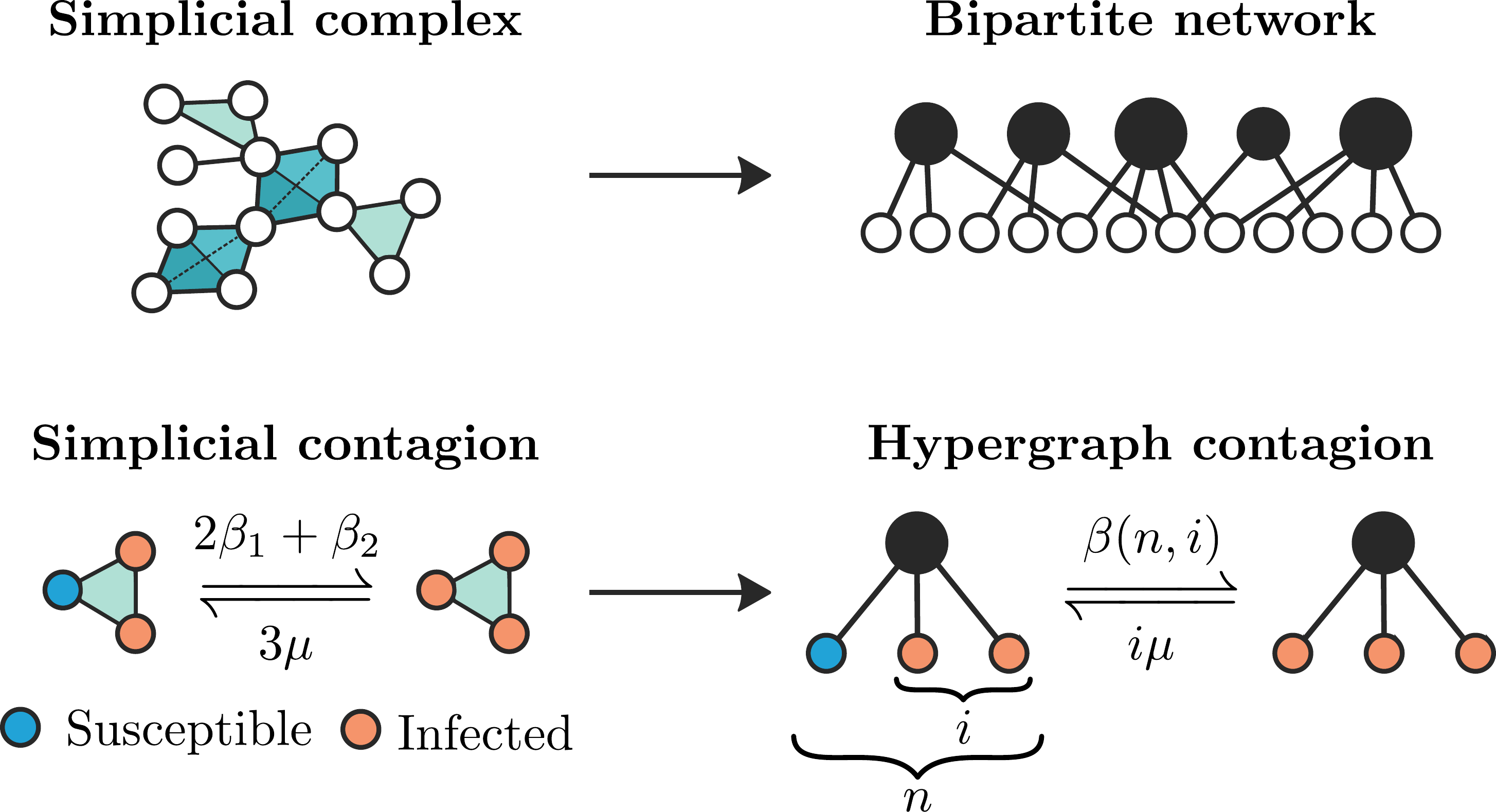}
\end{center}
\caption{{\bf Mapping of the simplicial contagion model to a hypergraph contagion.} We use a bipartite representation, where nodes (white circles) belong to \textit{groups} (black circles). A facet of dimension $n-1$ is mapped onto a group of size $n$. In the simplicial contagion model, contributions from higher-order interactions are taken into account by additional transmission rates (e.g., $\beta_2$ for the 2-simplex) when all but one node of the simplex are infected \cite{iacopini2019simplicial}. With our hypergraph representation, infections within a group are simply modeled by a general infection function $\beta(n,i)$ that depends on both the size $n$ of the group and the number $i$ of infected nodes in the group (with $i \leq n$).}
\label{fig:simplagion_mapping}
\end{figure}

In the original version of the simplicial contagion model~\cite{iacopini2019simplicial}, the spreading process takes place on a simplicial complex [see Fig.~\ref{fig:simplagion_mapping}] and obeys the following rule.
If all nodes in a $d$-simplex are infected except a susceptible one, this remaining node gets infected at a rate $\beta_d$, and also receives contributions from the lower-dimensional simplices included in the $d$-simplex with rates respectively equal to $\beta_{d-1},\dots,\beta_1$.
For instance, in Fig.~\ref{fig:simplagion_mapping}, two of the three nodes composing a 2-simplex are infected, hence the susceptible node gets infected at rate $2\beta_1 + \beta_2$, considering the contributions both from the two edges and from the ``triangle''.

A simplicial complex is a specific type of hypergraph, and thus we can always map the former on the latter---note that the reverse direction is not always possible.
To do so, each facet---a simplex that is not a face of any larger simplex---is represented by a single hyperedge (group).
In this paper, we relax the requirement of the simplicial complex in favor of a more general hypergraph structure.
We make use of the bipartite representation of hypergraphs~\cite{torres2020why,battiston2020networks}, in which the two sets of nodes of the bipartite graph correspond respectively to the sets of nodes and groups of the original hypergraph, as illustrated in Fig.~\ref{fig:simplagion_mapping}.

The \textit{size} $n$ of a group corresponds to the number of nodes belonging to this group, and it is therefore equivalent to the {\em hyperedge cardinality}.
Note that a $d$-simplex consists of $d+1$ nodes and is therefore mapped onto a group of size $n = d+1$.
Similarly, the \textit{membership} $m$ of a node, 
the {\em node hyperdegree}, corresponds to the number of groups to which it belongs, regardless of their size.

The hypergraph contagion model is defined as follows: for a group of size $n$, where $i \leq n$ members are infected, each of the $n-i$ susceptible nodes gets infected at rate $\beta(n,i)$.
For susceptible nodes that belong to multiple groups, their total transition rate to the infected state is simply the sum of the infection rates associated with each group to which they belong---in other words, the infection processes are independent.
All infected nodes transition back to the susceptible state at the same constant recovery rate $\mu$.

Notice that the hypergraph contagion model allows  to represent any type of simplicial contagion.
For instance, in the simplicial contagion model case of Fig.~\ref{fig:simplagion_mapping}, we would use $\beta(3,2) = 2 \beta_1 + \beta_2$.
In fact, the description offered by the infection rate function $\beta(n,i)$ yields a variety of models more general than the original simplicial contagion---in which a function $\beta(i)$ would be sufficient to encode contributions from facets of any dimension.

In all our case studies, we will use an infection rate function of the form
\begin{align}
    \label{eq:power_law_infection}
    \beta(n,i) = \lambda i^\nu \;.
\end{align}
However, many results we derive hold true for a general infection rate function $\beta(n,i)$.
The parameter $\nu$ controls the \textit{nonlinearity} of the contagion.
A linear contagion is recovered by setting $\nu = 1$, which is equivalent to a standard SIS model on networks, where each group is a clique~\cite{hebert-dufresne2010propagation}.
We intentionally chose the infection rate function independent of $n$ to focus on the impact of a nonlinear dependence on $i$; it would be straightforward to generalize the results by considering $\beta(n,i) \mapsto \Lambda(n) i^\nu$.

The infection rate function in  Eq.~\eqref{eq:power_law_infection} is the simplest nonlinear generalization of standard epidemiological models, where $\beta(n,i) \propto i$.
Moreover, we can motivate the choice of exponents $\nu \neq 1$ in the context of social contagions, by comparing our approach to the original formulation of the simplicial contagion model.
A value of $\beta_2 > 0$ in Fig.~\ref{fig:simplagion_mapping} represents \textit{social reinforcement}~\cite{iacopini2019simplicial}, and to correctly map the infection rate for a triangle, we need to use an exponent $\nu > 1$ in our model.
Similarly, a value $\beta_2 < 0$ represents \textit{social inhibition}, and this case can be  obtained with an exponent $\nu < 1$.

Another motivation for the infection rate function at Eq.~\eqref{eq:power_law_infection} is a recent study that shows this general form emerges in the occurrence of heterogeneous temporal patterns~\cite{st-onge2021bursty}.
More specifically, if you consider that the participation time of nodes---representing individuals---to higher-order interactions is distributed according to a power law, and that individuals become infected according to a threshold mechanism based on the \textit{dose} received in the interaction, then the probability for a node to get infected in a group is $\propto i^\nu$, where $\nu$ is related to the temporal heterogeneity.
In the continuous time limit, one recovers the infection rate function defined at Eq.~\eqref{eq:power_law_infection}.

In Ref.~\cite{st-onge2021bursty}, the infection mechanism is motivated in the context of biological contagions, where the infective dose received could represent viral particles for instance, and the threshold would correspond to the minimal infective dose to develop a disease.
While such type of complex contagions are rarely used in the context of biological contagions, they could help explain certain observed phenomena, such as super-exponential spread for certain diseases~\cite{scarpino2016effect}.
Moreover, threshold models are very common in social contagions \cite{granovetter1978threshold,watts2002simple,dodds2004universal,centola2010spread}, thus Eq.~\eqref{eq:power_law_infection} could be interpreted as an effective mechanism of social spread accounting for heterogeneous temporal patterns.

\subsubsection{Group-based Approximate Master Equations}

To describe hypergraph contagions, we make use of group-based Approximate Master Equations \cite{hebert-dufresne2010propagation,osullivan2015mathematical,st-onge2021social,st-onge2021master}.
This means that we do not rely on specific hypergraph realizations.
Instead, we assume that the structure is drawn from a random hypergraph ensemble described by the distributions $p_n$, for the size $n$ of a group, and $g_m$, for the membership $m$ of a node.
Each of the $m$ membership stubs of a node is assigned uniformly at random to a group available spot.
Therefore, the membership $m$ of a node and the sizes of the groups to which it belongs are uncorrelated.

To track the evolution of a contagion process on this ensemble of hypergraphs, we define two sets of quantities: $s_m(t)$, representing the fraction of nodes with membership $m$ that are susceptible at time $t$ and $f_{n,i}(t)$, the fraction of groups of size $n$ having $i$ infected members at time $t$.
The last quantity can also be interpreted as a conditional probability (of observing $i$ infected nodes in a group of size $n$) satisfying the normalization condition $\sum_{i} f_{n,i} = 1$.

We further define two mean-field quantities.
First, let us take a random susceptible node. The mean-field infection rate resulting from a random group to which it belongs is defined as
\begin{align}
    \label{eq:mean-field1}
    r(t) = \frac{\sum_{n,i}\beta(n,i)(n-i)f_{n,i} p_n}{\sum_{n,i} (n-i) f_{n,i} p_n} \;.
\end{align}
Indeed, the joint distribution for the size $n$ and the number of infected nodes $i$ in this group is proportional to $(n-i)f_{n,i} p_n$, and we just average $\beta(n,i)$ over this distribution.

Second, let us randomly choose a susceptible node inside a group. The mean-field infection rate caused by all the \textit{external} groups to which the susceptible node belongs (excluding the one from which we picked the node) can be written as
\begin{align}
    \label{eq:mean-field2}
    \rho(t) = r(t)\frac{\sum_{m} m(m-1) s_m g_m}{\sum_{m} m s_m g_m} \;.
\end{align}
To obtain $\rho(t)$, we assume that infections coming from different groups are independent processes.
We multiply $r(t)$ with the mean \textit{excess} membership of a susceptible node, i.e., if we pick a susceptible node in a group, it is the expected number of \textit{other} groups to which it belongs.
Since the membership distribution of a susceptible node picked in a group is proportional to $m s_m(t) g_m$, we simply average $m-1$, its excess membership, over this distribution.

Using the definitions in Eqs.~\eqref{eq:mean-field1} and \eqref{eq:mean-field2} we can write the following system of AMEs
\begin{subequations}\label{eq:simplagion_ODE}
\begin{align}
    \d{s_m}{t} =& \mu(1 - s_m) - m r s_m  \;, \label{eq:simplagion_ODE_sm}\\ 
    \d{f_{n,i}}{t} =& \mu (i+1) f_{n,i+1} - \mu i f_{n,i} \; 
    \notag \\
                    &- (n-i)\br{\beta(n,i) + \rho}f_{n,i}\; \label{eq:simplagion_ODE_fni}\\
                    &+ (n-i+1)\br{\beta(n,i-1) + \rho}f_{n,i-1} \notag \;.
\end{align}
\end{subequations}
This system is composed of $\mathcal{O}(n_\mathrm{max}^2 + m_\mathrm{max})$ equations. 
It is approximate in the sense that the evolution of the fractions of infected nodes of membership $m$, $s_m$, are treated in a mean-field fashion (still considering dynamic correlations between pairs node-group), while the evolution equations of the $f_{n,i}$ are treated as master equations. In the right-hand side of Eq. (\ref{eq:simplagion_ODE_fni}), the first two terms are due to the recovery process, and the last two to the infection. The infection rate due to infected nodes \textit{inside} the group is exact, while the infection rates associated to \textit{external} groups (the terms making use of $\rho$) are treated in an approximate way.
Without loss of generality (up to a change of time scale) we set $\mu \equiv 1$ for the remainder of the document.

From Eq.~\eqref{eq:simplagion_ODE}, we can calculate the global prevalence
\begin{align}
    I(t) = \sum_m [1 - s_m(t)] g_m\;,
\end{align}
and the group prevalence
\begin{align}
    I_n(t) = \sum_i \frac{i}{n} f_{n,i}(t) \;,
\end{align}
which correspond to the average fraction of infected nodes in the whole system and within groups of size $n$ respectively.

In the stationary state, we obtain the following self-consistent relations:
\begin{subequations}\label{eq:stationary_state}
\begin{align}
    s_m^* =& \frac{1}{1 + m r^*} \;, \label{eq:stationary_state_sm}\\
    (i+1) f_{n,i+1}^* =& \cbrace{i + (n-i)\br{\beta(n,i) + \rho^*}}f_{n,i}^* \notag \\
                       &- (n-i+1)\br{\beta(n,i-1) + \rho^*}f_{n,i-1}^* \;. \label{eq:stationary_state_fni}
\end{align}
\end{subequations}
The latter equation can be simplified by noting that $f_{n,i}^*$ must respect \textit{detailed balance} in the stationary state, i.e.,
\begin{align*}
    (i+1) f_{n,i+1}^* &= (n-i)\br{\beta(n,i) + \rho^*}f_{n,i}^* \;,
\end{align*}
expressing that the probability to decrease the number of infectious nodes in a group of size $n$ from $i+1$ to $i$ by a recovery process is equal to the probability of the opposite change (from $i$ to $i+1$ infectious nodes) obtained through a contagion event.
We thus finally obtain
\begin{align}
    \label{eq:fni_explicit}
    f_{n,i}^* = f_{n,0}^* \frac{n!}{(n-i)!i!} \prod_{j=0}^{i-1} [ \beta(n,j) + \rho^*] \quad \forall \; i \in \lbrace 1, \dots, n \rbrace\;,
\end{align}
where $f_{n,0} = 1-\sum_{i>0} f_{n,i}$.

\begin{figure*}[tb]
\begin{center}
    \includegraphics[width = \linewidth ]{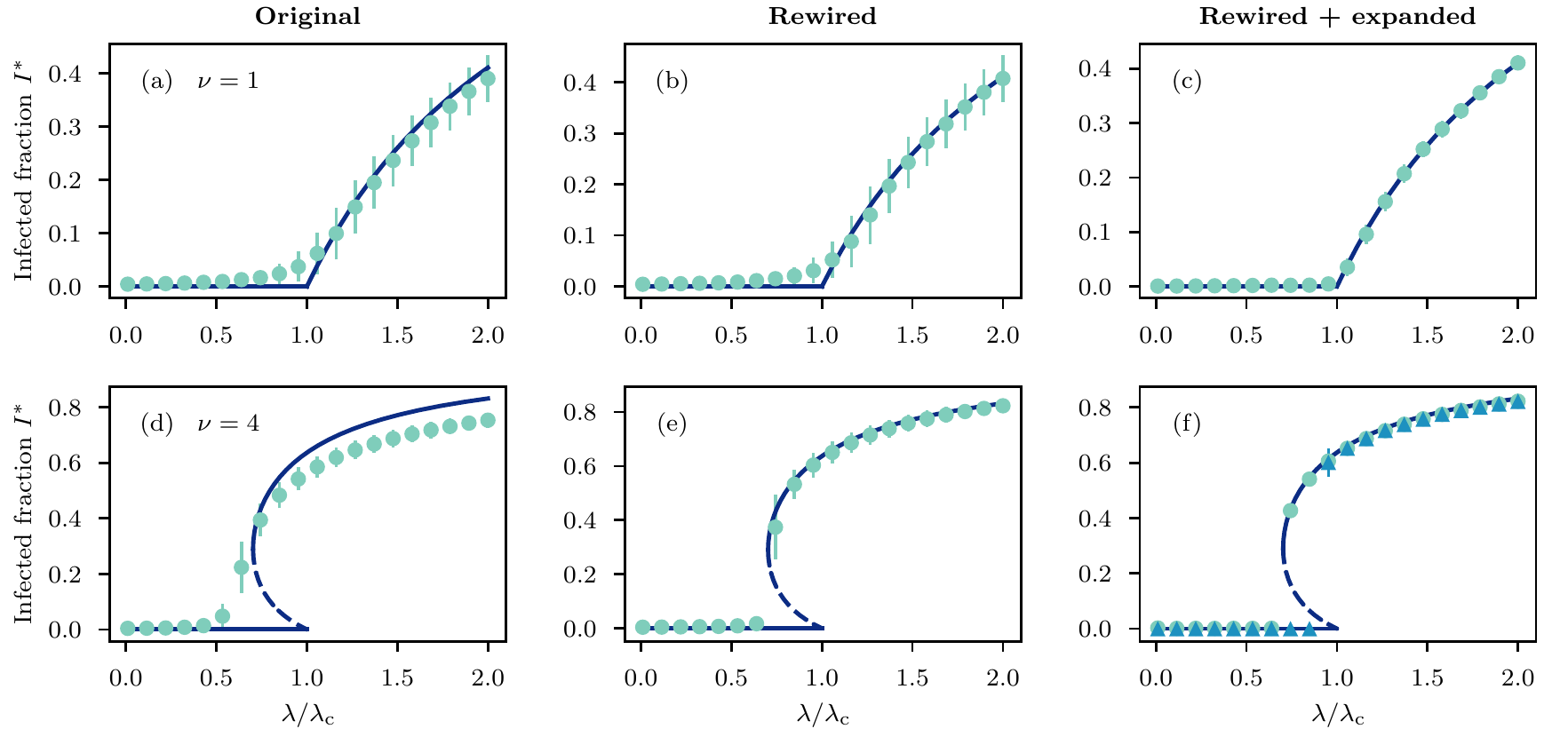}
\end{center}
\caption{{\bf Fraction $\boldsymbol{I^*}$ of infectious nodes in the stationary state of a contagion model on a hypergraph constructed from high-resolution face-to-face contact data from a primary school in Lyon (see Methods Sec.~\ref{app:datasets}).}
The hypergraph contains $242$ nodes and $1188$ groups.
Both the membership and the group size distributions are homogeneous, with $\langle m \rangle \approx 11.79$, $\langle n \rangle \approx 2.40$, $m_\mathrm{max} = 32$ and $n_\mathrm{max} = 5$.
We compare the results of Monte Carlo simulations (symbols; see Methods Sec.~\ref{app:simulation}) with the predictions of our approach (solid and dashed lines for stable and unstable solutions respectively). We rescale $\lambda$ with the invasion threshold $\lambda_\mathrm{c}$, which is computed using Eq.~\eqref{eq:invasion_threshold} in Sec.~\ref{sec:phase-transition}.
The symbols represent the average infected fraction measured over long runs and averaged over randomized hypergraphs when this applies. The error bars (sometimes too small to be seen) correspond to one standard deviation.
The green circles are obtained with the quasistationary-state method, starting with a large fraction of infected nodes $(I = 0.8)$ to sample the upper branch of the hysteresis loop when the phase transition is discontinuous.
(a),(d) We use the original hypergraph. (b),(e) We use 10 uniformly randomized versions of the original hypergraph. (c),(f) We use 10 uniformly randomized versions of the original hypergraph with its size expanded by a factor 10 (see Methods Sec.~\ref{app:data-augmentation}).
(f) The blue triangles are obtained by ordinary simulations, starting with a small fraction of infected nodes $(I = 0.02)$ to sample the lower branch of the hysteresis loop. We only do it for the expanded hypergraphs because finite-size effects makes unreliable the estimation of the lower branch with data of small size.
}
\label{fig:simulation_sociopatterns}
\end{figure*}

\begin{figure*}[tb]
\begin{center}
    \includegraphics[width = \linewidth ]{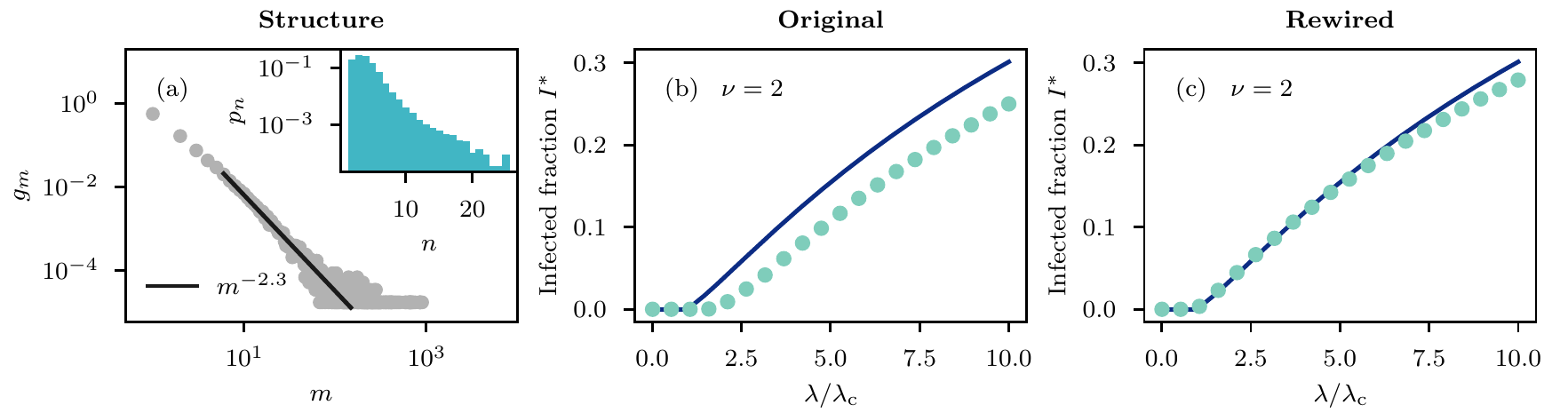}
\end{center}
\caption{{\bf Fraction $\boldsymbol{I^*}$ of infectious nodes in the stationary state of a contagion model on a hypergraph constructed from coauthorship data in computer science (see Methods Sec.~\ref{app:datasets}).}
The hypergraph has been obtained by using a breadth-first search to extract a sub-hypergraph of $57\,501$ nodes and $55\,204$ groups; the original data contained more than $10^6$ nodes and groups.
(a) The membership distribution is heterogeneous, approximately of the form $g_m \sim m^{-\gamma_m}$, while the group size distribution is more homogeneous, with $\langle m \rangle \approx 3.75$, $\langle n \rangle \approx 3.90$, $m_\mathrm{max} = 903$ and $n_\mathrm{max} = 25$.
(b),(c) We compare the results of Monte Carlo simulations (circle markers; see Methods Sec.~\ref{app:simulation}) with the predictions of our approach (solid lines).
We rescale $\lambda$ with the invasion threshold $\lambda_\mathrm{c}$, which is computed using Eq.~\eqref{eq:invasion_threshold} in Sec.~\ref{sec:phase-transition}.
The infected fraction $I^*$ has been obtained as averages over time with long runs (and averaging over randomized hypergraphs when this applies). The error bars (too small to be seen) correspond to one standard deviation.
We used ordinary simulations, starting with $I = 0.8$.
(b) We use the original hypergraph. (c) We use 10 uniformly randomized versions of the original hypergraph.}
\label{fig:simulation_coauthorship}
\end{figure*}

\subsubsection{Comparison with simulations}
\label{sec:simulation}


We provide a comparison of our approach with Monte Carlo simulations (see Methods Sec.~\ref{app:simulation}). To do this, we consider empirical higher-order structures constructed from real-world data, and their randomized counterparts. Details on the data collection and aggregation are provided in Methods Sec.~\ref{app:datasets}.

The motivation for this \textit{a priori} validation is threefold: first, it allows us to illustrate the validity and accuracy of our analytical framework when our assumptions are met---namely when the structure is drawn from an ensemble of uncorrelated hypergraphs with fixed $g_m$ and $p_n$.
Second, because of the excellent agreement with simulations for random hypergraphs, we omit further comparison with Monte Carlo simulations for the many results we present in the following sections.
Finally, it showcases the possible sources of discrepancies---and how our results could vary---when considering real datasets.

In Fig.~\ref{fig:simulation_sociopatterns}, we show the
phase diagram, i.e., the fraction $I^*$ of infectious nodes in the stationary state, of contagion dynamics on hypergraphs that encode higher-order social (face-to-face) interactions between individuals from a primary school in Lyon (see Methods Sec.~\ref{app:datasets} for details). Both the membership and group size distributions are homogeneous for this dataset. We considered linear contagion $(\nu = 1)$, equivalent to the standard SIS model, and superlinear contagion $(\nu = 4)$. In both cases, our analytical formalism (continuous lines) agrees quite well with the Monte Carlo simulations (symbols) on the original (empirical) hypergraph [Figs.~\ref{fig:simulation_sociopatterns}(a) and \ref{fig:simulation_sociopatterns}(d)]. The main source of errors for the observed discrepancy can be attributed to \textit{structural} correlations, which do not appear to affect the threshold values, but reduce the stationary prevalence. Indeed, in Figs.~\ref{fig:simulation_sociopatterns}(b) and \ref{fig:simulation_sociopatterns}(e), the agreement improves by randomizing the hypergraph while preserving the memberships and the group sizes. The remaining discrepancies are due to the fact that simulations are affected by finite-size effects, while our formalism assumes an infinite size system: the agreement becomes almost perfect in Figs.~\ref{fig:simulation_sociopatterns}(c) and \ref{fig:simulation_sociopatterns}(f) by additionally increasing the size of the hypergraph by a factor $10$.  

Let us remark that the error is more important for \mbox{$\nu = 4$} on the original hypergraph [Fig.~\ref{fig:simulation_sociopatterns}(d)], which suggests that structural correlations have a greater effect on nonlinear contagions. In the Supplementary Information, we show how to generalize our framework to account for structural correlations.

We also considered a completely different dataset, which represents co-authorship relations in computer science publications, obtained from major journals and proceedings in the field. The resulting hypergraph is considerably larger than the previous one, and it also presents a very different structure (see Methods Sec.~\ref{app:datasets}). The results are shown in Fig.~\ref{fig:simulation_coauthorship}, where we plot the same phase diagram curves as in Fig. \ref{fig:simulation_sociopatterns}, using a superlinear contagion $(\nu = 2)$. In this case, however, the membership distribution is heterogeneous [Fig.~\ref{fig:simulation_coauthorship}(a)], approximately of the form $g_m \sim m^{-\gamma_m}$ with $\gamma_m \approx 2.3$, while the group size distribution is more homogeneous [inset of Fig.~\ref{fig:simulation_coauthorship}(a)], but still extends to rather large values with $n_\mathrm{max} = 25$.
By comparing the results for the original hypergraph [Fig.~\ref{fig:simulation_coauthorship}(b)] against those for a randomized ensemble [Fig.~\ref{fig:simulation_coauthorship}(c)], we see that structural correlations account for the major part of the discrepancies between simulations and theory, affecting both the invasion threshold and the stationary prevalence.
However, Fig.~\ref{fig:simulation_coauthorship}(c) shows that structural correlations are not the only source of errors at high prevalence.

Let us recall that our formalism correctly captures the \textit{dynamical} correlations within a group through a master equation description [Eq.~\eqref{eq:simplagion_ODE_fni}] of the possible states, but it does not capture the dynamical correlations around nodes, since we use a heterogeneous mean-field description [Eq.~\eqref{eq:simplagion_ODE_sm}].
These correlations become especially important in the presence of hubs with a large membership, which is the case for the hypergraph considered in Fig.~\ref{fig:simulation_coauthorship}.
In fact, when we use the same group size distribution as in Fig.~\ref{fig:simulation_coauthorship}, but with a more homogeneous membership distribution, the discrepancy at high prevalence disappears (see the Supplementary Information).

\subsection{Phase transition}
\label{sec:phase-transition}

\begin{figure}
\begin{center}
    \includegraphics[width = \linewidth]{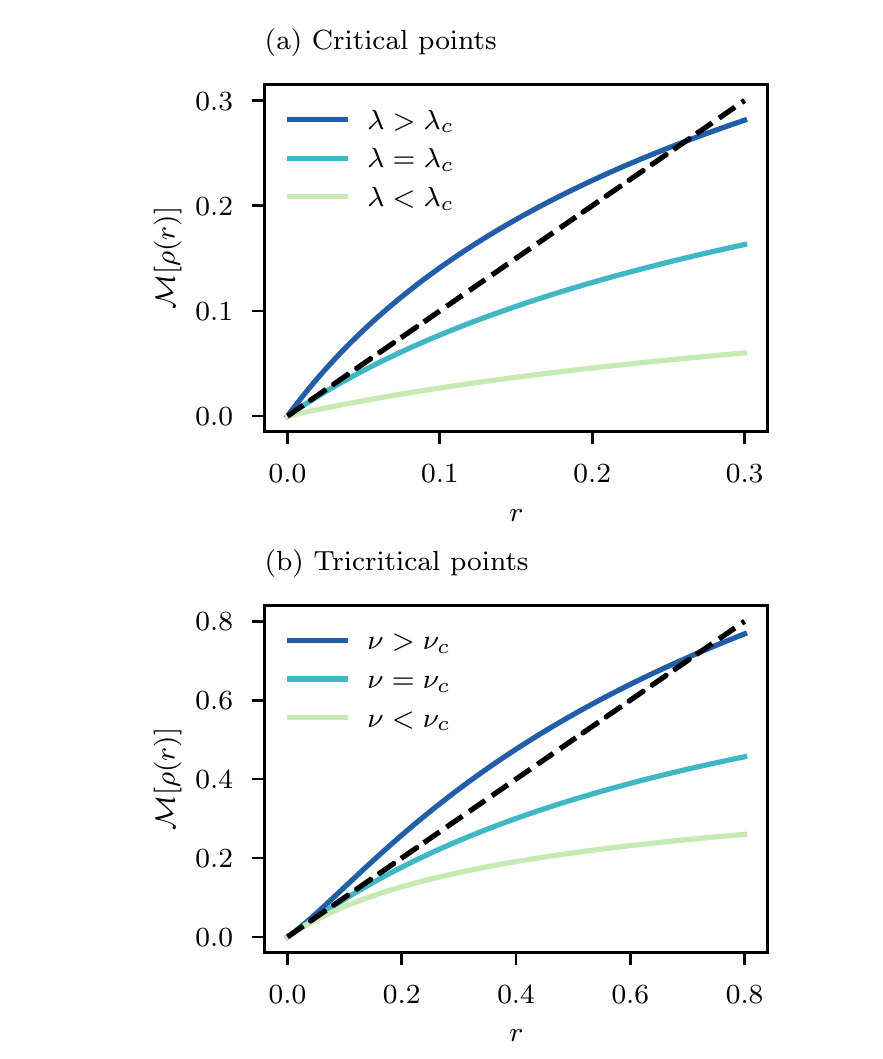}
\end{center}
\caption{{\bf Critical behavior of the function $\boldsymbol{\mathcal{M}[\rho(r)]}$.} Each intersection with the dashed line represents a stationary solution of Eq.~\eqref{eq:stationary_state}. Results refer to the infection rate function $\beta(n,i) = \lambda i^\nu$ and a hypergraph with $g_m =(\delta_{m,2} + \delta_{m,3})/2$ and $p_n = (\delta_{n,4} + \delta_{n,5})/2$. (a) We fix $\nu = 1$ and vary $\lambda$. For $\lambda > \lambda_\mathrm{c}$, a non-trivial solution emerges. (b) We fix $\lambda = \lambda_\mathrm{c}$ and vary $\nu$. Note that the slope of all solid lines is 1 at the origin. A nonlinear exponent $\nu > \nu_\mathrm{c}$ is associated with a discontinuous phase transition, since $\partial_r^2\mathcal{M} > 0$ in this case.}
\label{fig:map}
\end{figure}

\begin{figure*}
\begin{center}
    \includegraphics[width = \linewidth ]{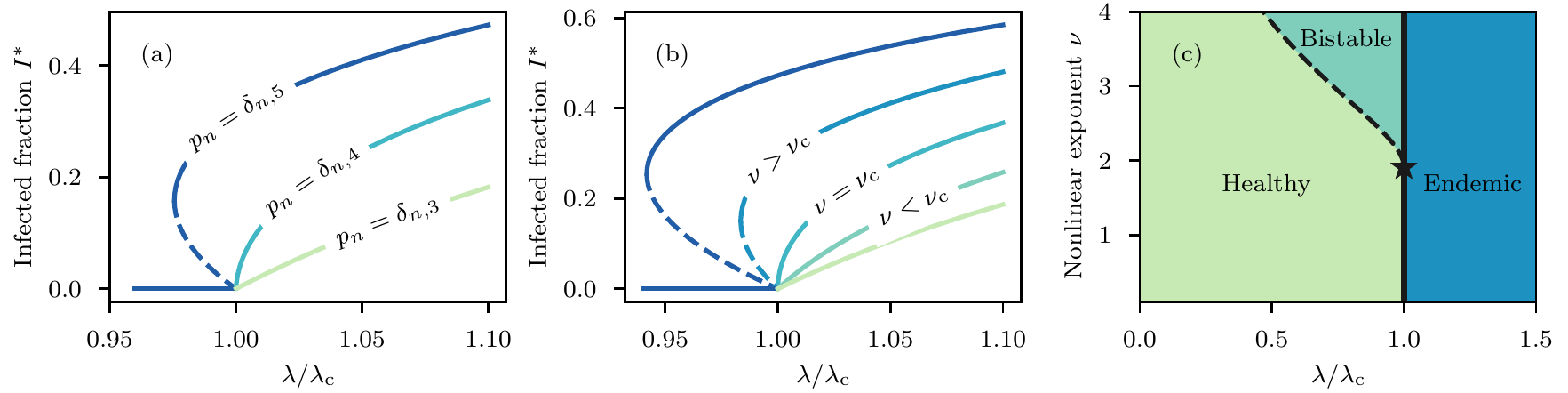}
\end{center}
\caption{{\bf Phase transition in regular hypergraphs (hypergraphs with fixed $\boldsymbol{m}$ and $\boldsymbol{n}$).} The infection rate function is $\beta(n,i) = \lambda i^\nu$. Solid and dashed lines in (a) and (b) represent stable and unstable solutions respectively for the stationary fraction of infected nodes. (a) $\nu = 1.8$, but different structures were used, with a node always having the same number of different neighbors. From bottom to top, we have $(g_m,p_n) = (\delta_{m,6},\delta_{n,3})$, $(\delta_{m,4},\delta_{n,4})$ and $(\delta_{m,3},\delta_{n,5})$. Note that for the middle curve ($\delta_{m,4},\delta_{n,4}$), $\nu_\mathrm{c} \approx 1.81$. (b) The hypergraph structure is fixed with $g_m = \delta_{m,3}$ and $p_n = \delta_{n,4}$. Values of $\nu \in \cbrace{1.5,1.7, \nu_\mathrm{c}, 2.1, 2.3}$ (bottom to top curves) were used, with $\nu_\mathrm{c} \approx 1.91$. (c) Phase diagram for the same hypergraph as in (b). The dashed critical line ($\lambda = \lambda_\mathrm{p}$) and the solid critical line ($\lambda = \lambda_\mathrm{c}$) coalesce at the tricritical point $(\lambda_\mathrm{c},\nu_\mathrm{c})$ indicated by the star marker.}
\label{fig:bifurcation_hom}
\end{figure*}

In this section, we unveil the important role of influential groups on the phase transition of hypergraph contagions.
We first derive general expression for critical points, marking the limit of the domain of validity of a stationary solution.
Secondly, we obtain an expression for tricritical points, indicating when the phase transition switches from continuous to discontinuous, and a bistable regime appears.
These results are valid for any infection rate function such that $\beta(n,i) > 0$ for all $i > 0$ (see Supplementary Information for a consideration of threshold models).

We then concentrate our study on the infection rate function $\beta(n,i) = \lambda i^\nu$.
This allows us to define important threshold values, the \textit{invasion threshold} $\lambda_\mathrm{c}$ above which the disease-free solution $I^* = 0$ is unstable, and the \textit{bistability threshold} $\nu_\mathrm{c}$, the smallest nonlinear exponent allowing for a discontinuous phase transition.
In particular, we will show how heterogeneous membership and group size distributions alter these thresholds, especially in the presence of a mesoscopic localization  driven by influential groups.

In the following, we assume that the system has reached the stationary state and we drop the asterisk to simplify the notation throughout this section.

\subsubsection{Critical points and the invasion threshold}

Equations~\eqref{eq:stationary_state_sm} and \eqref{eq:fni_explicit} imply that each $s_m$ and $f_{n,i}$ can be written in terms of $r$ and $\rho$.
In turn, $r$ and $\rho$ can be written in terms of $s_m$ and $f_{n,i}$ through Eqs.~\eqref{eq:mean-field1} and \eqref{eq:mean-field2}, which means the stationary solutions are determined by a set of self-consistent equations.

Satisfying all self-consistent equations means we can reexpress all quantities ($s_m,f_{n,i},\rho,r$) as functions of a single mean-field quantity, which we choose to be $r$.
$s_m(r)$ is given by Eq.~\eqref{eq:stationary_state_sm}, and $\rho(r)$ is given by Eq.~\eqref{eq:mean-field2}, which is rewritten as
\begin{align}
    \label{eq:mean-field2_stationary}
    \rho(r) = r  \frac{\sum_{m} m(m-1) s_m(r) g_m}{\sum_{m} m s_m(r) g_m} \;.
\end{align}
$f_{n,i}$ is more simply written as a composite function, $f_{n,i}[\rho(r)]$, given by Eq.~\eqref{eq:fni_explicit}.
Finally, $r$ itself must satisfy Eq.~\eqref{eq:mean-field1}, which we can write as $r = \mathcal{M}[\rho(r)]$ where
\begin{align}
    \label{eq:map}
    \mathcal{M}(\rho) = \frac{\sum_{n,i}\beta(n,i)(n-i)f_{n,i}(\rho) p_n}{\sum_{n,i} (n-i) f_{n,i}(\rho) p_n}  \;.
\end{align}

In Fig.~\ref{fig:map}, the stationary solutions thus correspond to the intersections of $\mathcal{M}[\rho(r)]$ (solid lines) and $r$ (dashed line).
We see that $r = 0$ is always a solution, while the solution $r > 0$ only exists for certain values of the parameter $\lambda$. This indicates the presence of a critical point.

Critical points mark the limit of the domain of validity of a solution to the equation \mbox{$r = \mathcal{M}[\rho(r)]$}. They arise when $r$ is tangent to $\mathcal{M}[\rho(r)]$, which implies
\begin{align}
    \d{\mathcal{M}}{r} &= \d{\mathcal{M}}{\rho} \d{\rho}{r} = 1 \label{eq:threshold_non_explicit}\;,
\end{align}
as can be seen in Fig.~\ref{fig:map}(a), where $\mathcal{M}[\rho(r)]$ is tangent to $r$ at the point $r = 0$ for some value $\lambda = \lambda_\mathrm{c}$.

When the tangent point $r > 0$, Eq.~\eqref{eq:threshold_non_explicit} needs to be solved numerically.
However when the solution $r \to 0$, we are able to obtain an analytical expression.
In this limit, we expect to have $s_m \to 1$, $\rho \to 0$ and \mbox{$f_{n,i} \to \delta_{i,0} \,\forall n$}, i.e., all nodes are susceptible.
Therefore, from Eq.~\eqref{eq:mean-field2_stationary} we have
\begin{align}
    \label{eq:first_d}
    \left.\d{\rho}{r}\right|_{r \to 0} = \frac{\avg{m(m-1)}}{\avg{m}} \;,
\end{align}
where $\avg{\cdots}$ stands for the expectation value with respect either to $g_m$ or $p_n$.
From Eq.~\eqref{eq:map} (and using the fact that $\beta(n,0) \equiv 0$) we also obtain
\begin{align}
    \label{eq:second_d}
    \left.\d{\mathcal{M}}{\rho}\right|_{\rho \to 0} = \frac{1}{\avg{n}} \sum_{n,i} \beta(n,i) (n-i) p_n \underbrace{\left.\d{f_{n,i}}{\rho}\right|_{\rho \to 0}}_{\displaystyle{h_{n,i}}}  \;.
\end{align}
To evaluate $h_{n,i}$, we apply the derivative $\mathrm{d}/\mathrm{d}\rho $ to Eq.~\eqref{eq:fni_explicit}.
We obtain $h_{n,1} = n$ and
\begin{align}
    \label{eq:hni_explicit}
    h_{n,i} = \frac{n! \prod_{j = 1}^{i-1} \beta(n,j)}{(n-i)! \; i!} \quad \forall \; i \in \lbrace 2, \dots, n \rbrace \;,
\end{align}
Also, $\sum_{i} h_{n,i} \equiv 0$ as $\sum_i f_{n,i}=1$, hence
\begin{align*}
    h_{n,0} = - \sum_{i > 0} h_{n,i} \;.
\end{align*}

Combining Eqs~\eqref{eq:first_d}, \eqref{eq:second_d} and \eqref{eq:hni_explicit} with Eq.~\eqref{eq:threshold_non_explicit}, we obtain the following implicit expression for critical points in the limit $r \to 0$
\begin{align}
    \label{eq:invasion_threshold}
    \frac{\avg{m(m-1)}}{\avg{m} \avg{n}} \left \langle \sum_{i > 0}  \frac{ \;n! }{(n-i-1)! \; i!} \prod_{j = 1}^{i} \beta(n,j) \right \rangle = 1 \;.
\end{align}

For the infection rate function $\beta(n,i) = \lambda i^\nu$, Eq.~\eqref{eq:invasion_threshold} defines the \textit{invasion} threshold $\lambda_\mathrm{c}$, i.e., the critical value of $\lambda$ marking the limit of the validity for a solution $r \to 0$. This solution is not always stable, but we always have that the trivial solution $r = 0$, corresponding to $I = 0$, is unstable for all $\lambda > \lambda_\mathrm{c}$.
This is illustrated in Fig.~\ref{fig:bifurcation_hom}(a) and \ref{fig:bifurcation_hom}(b) for regular hypergraph structures with fixed membership and group size.
The invasion threshold depends on both the structure $(g_m,p_n)$ and the nonlinear exponent $\nu$.
The resulting phase diagram in the $(\lambda, \nu)$ plane is shown in Fig.~\ref{fig:bifurcation_hom}(c). The invasion threshold spans the critical line (solid line) $\lambda = \lambda_\mathrm{c}$ in the phase diagram.

The dashed critical line $\lambda = \lambda_\mathrm{p}$ is associated with the limit of validity of a solution where $\mathcal{M}[\rho(r)]$ is tangent to $r$ for some $r > 0$---it is thus solved numerically.
We call $\lambda_\mathrm{p}$ the \textit{persistence} threshold as it is the smallest value of $\lambda$ such that a non-trivial solution is locally stable.
For continuous phase transitions, $\lambda_\mathrm{c} = \lambda_\mathrm{p}$, but for discontinuous phase transitions, $\lambda_\mathrm{p} < \lambda_\mathrm{c}$.

\subsubsection{Tricritical points and the bistability threshold}

Depending on the structure and the form of $\beta(n,i)$, we can have a continuous or a discontinuous phase transition, as can be seen in Fig.~\ref{fig:bifurcation_hom}.
When the phase transition is continuous, we have possibly two solutions for the stationary fraction of infected nodes, $I_1 = 0$ and $I_2 > 0$.
When $I_2$ exists (for instance when $\lambda > \lambda_\mathrm{c}$), $I_1$ is unstable.

When the phase transition is discontinuous, we have typically three solutions, $I_1 = 0$ and $0 < I_2 < I_3$.
In the bistable regime [for instance when $\lambda \in (\lambda_\mathrm{p},\lambda_\mathrm{c})$], all three solutions coexist, $I_2$ is unstable, and $I_1$ and $I_3$ are locally stable.
In the endemic regime [for instance when $\lambda \geq \lambda_\mathrm{c}$], only $I_1$ and $I_3$ exist, and only $I_3$ is locally stable.

We are interested in knowing when the phase transition changes from continuous to discontinuous.
In Fig.~\ref{fig:bifurcation_hom}(c), the bistable regime starts at a tricritical point (star marker), where two critical lines meet.
For the infection rate function $\beta(n,i) = \lambda i^\nu$ and a fixed hypergraph structure, the tricritical point happens at $(\lambda,\nu) = (\lambda_\mathrm{c},\nu_\mathrm{c})$, where $\nu_\mathrm{c}$ is what we call the \textit{bistability} threshold, since a bistable regime only exists for $\nu > \nu_\mathrm{c}$.

To get some insights on the properties of tricritical points, we show in Fig.~\ref{fig:map}(b) the function $\mathcal{M}[\rho(r)]$ at \mbox{$\lambda = \lambda_\mathrm{c}$} and for values of $\nu$ below, at and above the bistability threshold.
For $\nu < \nu_\mathrm{c}$, we have $\partial_r^2 \mathcal{M}|_{r\to 0} < 0$ and $I_3$ does not exist.
For $\nu > \nu_\mathrm{c}$, we have $\partial_r^2 \mathcal{M}|_{r\to 0} > 0$ and there exists a solution $I_3 > 0$.
At the tricritical point, $I_3 = I_2 \to 0$, hence the non-trivial solution is degenerate, which is possible only if
\begin{align*}
    \left.\dd{\mathcal{M}}{r} \right|_{r \to 0} = \left.\br{\d{\mathcal{M}}{\rho} \dd{\rho}{r} + \pr{\d{\rho}{r}}^2 \dd{\mathcal{M}}{\rho} } \right|_{r \to 0} = 0 \;.
\end{align*}
Since a tricritical point is also a critical point, from Eq.~\eqref{eq:threshold_non_explicit} $\mathrm{d} \mathcal{M}/ \mathrm{d} \rho = (\mathrm{d} \rho / \mathrm{d} r)^{-1}$, so the condition can be rewritten as
\begin{align*}
    \left.\br{\dd{\rho}{r} + \pr{\d{\rho}{r} }^3 \dd{\mathcal{M}}{\rho} } \right|_{r \to 0} = 0 \;.
\end{align*}

The derivatives on $\rho$ with respect to $r$ at a critical point where $r \to 0$ can be easily evaluated, and the condition now becomes
\begin{align}\label{eq:dev_rho_lambdac}
    2\pr{\frac{\avg{m^2}^2}{\avg{m}^2} - \frac{\avg{m^3}}{\avg{m}}} + \pr{\frac{\avg{m(m-1)}}{\avg{m}}}^3 \left. \dd{\mathcal{M}}{\rho} \right|_{\rho \to 0} = 0 \;.
\end{align}
To evaluate the last term of Eq.~\eqref{eq:dev_rho_lambdac}, let us rewrite
\begin{align*}
    \mathcal{M}(\rho) = \frac{\sum_{n,i}\beta(n,i)(n-i)f_{n,i}(\rho) p_n}{\sum_{n,i} (n-i) f_{n,i}(\rho) p_n} \equiv \frac{u(\rho)}{v(\rho)} \;.
\end{align*}
In the limit $\rho \to 0$, $f_{n,i} \to \delta_{i,0}$, which implies  $u(\rho) \to 0$ and \mbox{$v(\rho) \to \avg{n}$}, therefore 
\begin{align}
    \left.\dd{\mathcal{M}}{\rho} \right|_{\rho \to 0} = \left. \frac{1}{\avg{n}} \dd{u}{\rho}\right|_{\rho \to 0} - \left. \frac{2}{\avg{n}^2}\d{u}{\rho} \d{v}{\rho} \right|_{\rho \to 0} \;.
\end{align}
First order derivatives can be evaluated using
\begin{subequations}
\begin{align}
    \left. \d{u}{\rho}  \right|_{\rho \to 0} &= \sum_{n,i} \beta(n,i) (n-i) p_n h_{n,i} \;,\\
    \left. \d{v}{\rho}  \right|_{\rho \to 0} &= \sum_{n,i} (n-i) p_n h_{n,i} \;.
\end{align}
\end{subequations}
For the second-order derivative, let us define \mbox{$l_{n,i} \equiv \mathrm{d}^2 f_{n,i} / \mathrm{d} \rho^2 |_{\rho \to 0}$}, so that we can write
\begin{align}
    \left. \dd{u}{\rho}\right|_{r \to 0} &= \sum_{n,i} \beta(n,i) (n-i) p_n l_{n,i} \;.
\end{align}
Finally, we can apply the second order derivative to Eq.~\eqref{eq:stationary_state_fni} to obtain the recurrence relation
\begin{align}
    (i+1) l_{n,i+1} =& 2\br{(n-i) h_{n,i} - (n-i+1) h_{n,i-1}} \;, \notag \\
                     &+ \br{i + (n-i)\beta(n,i)}l_{n,i} \;, \\
                     &- (n-i+1)\beta(n,i-1) l_{n,i-1} \;. \notag
\end{align}
Again, $l_{n,0} = - \sum_{i>0} l_{n,i}$ by definition.

Even though it is possible to express the $l_{n,i}$ in an explicit form, the expression does not give us more intuition, and it is simpler to calculate the $l_{n,i}$ using the recurrence equation just given. After rewriting \mbox{$\mathrm{d}^2 \mathcal{M}/\mathrm{d} \rho^2|_{\rho \to 0} \equiv F[p_n, \beta(n,i)]$}, tricritical points are obtained by solving the equation
\begin{align}
    \label{eq:bistability_threshold}
    \frac{\avg{m^2}^2}{\avg{m}^2} - \frac{\avg{m^3}}{\avg{m}} + \frac{\avg{m(m-1)}^3}{2\avg{m}^3} F[p_n, \beta(n,i)] = 0 \;.
\end{align}

Tricritical points result from an intricate relation between the structure ($g_m, p_n$) and the infection rate $\beta(n,i)$.
Fig.~\ref{fig:bifurcation_hom} shows that either changing the structure [Fig.~\ref{fig:bifurcation_hom}(a)] or the shape of the infection rate function [Fig.~\ref{fig:bifurcation_hom}(b)-(c)] can lead to a change of behavior, from a continuous phase transition to a discontinuous one with a bistable regime.

A first hypothesis we can make from these simple examples is that more nonlinear infection rates (larger $\nu$) and larger groups promote bistability.
However, we will see that this intuition does not hold in general for heterogeneous structures due to the onset of mesoscopic localization.

\subsubsection{Heterogeneous memberships}

\begin{figure*}[tb]
\begin{center}
    \includegraphics[width = \linewidth ]{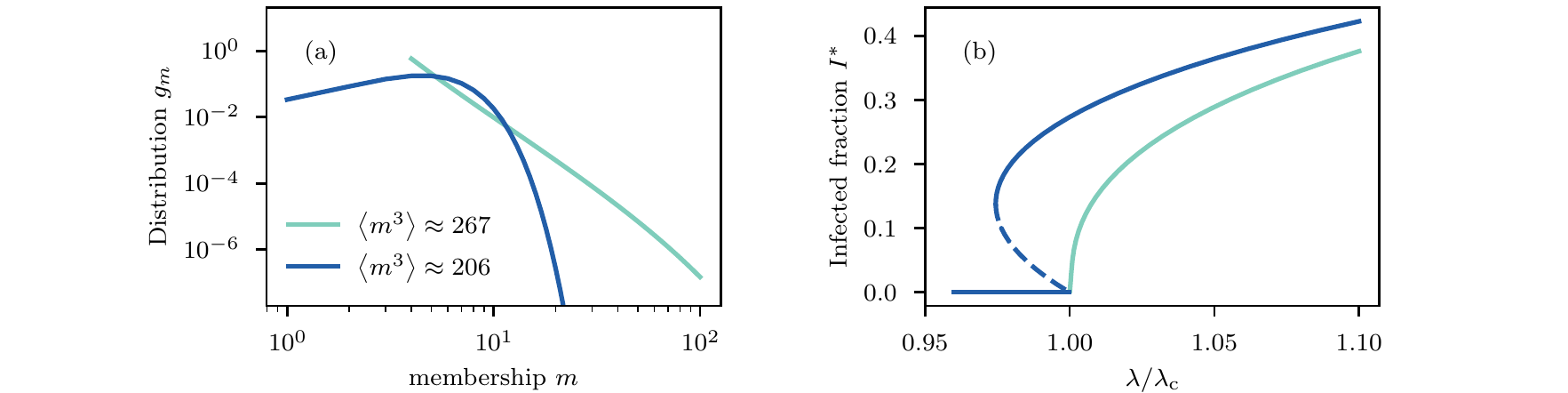}
\end{center}
\caption{{\bf Impact of the third moment of the membership distribution on the phase transition.} The group size distribution is fixed to $p_n = \delta_{n,3}$. (a) Membership distributions with the same first two moments ($\avg{m} \approx 5.03$ and $\avg{m^2}\approx 30.2$) but with a different third moment $\avg{m^3}$. The distribution with higher $\avg{m^3}$ is of the form $g_m \sim (a+m)^{-b}e^{-m/c}$ with $(a,b,c) = (-0.8,3.72,40)$ and ranges from $m_\mathrm{min} = 4$ to $m_\mathrm{max} = 100$. The one with lower $\avg{m^3}$ is Poisson distributed $g_m \sim a^m e^{-a} /m!$ with $a = 5$ and ranges from $m_\mathrm{min} = 1$ to $m_\mathrm{max} = 100$.  (b) Non-trivial solutions to Eq.~\eqref{eq:stationary_state} for the corresponding membership distribution. Solid and dashed lines represent stable and unstable solutions, respectively. The infection rate function considered is $\beta(n,i) = \lambda i^\nu$ with $\nu = 2.8$.}
\label{fig:thirdmoment}
\end{figure*}

\begin{figure*}[tb]
\begin{center}
    \includegraphics[width = \linewidth ]{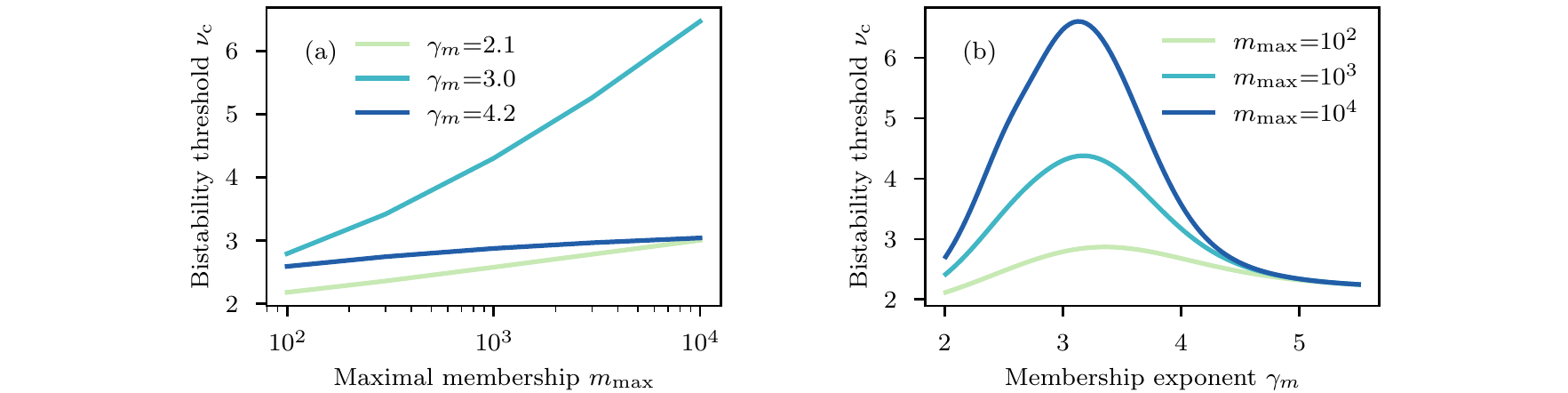}
\end{center}
\caption{{\bf Impact of heterogeneous memberships on the bistability threshold.} We considered hypergraphs with power-law membership distributions $g_m \sim m^{-\gamma_m}$ with various exponents $\gamma_m$, different maximal values $m_\mathrm{max}$, a minimal value $m_\mathrm{min} =2$, and a regular group size distribution $p_n = \delta_{n,4}$. We obtain the bistability threshold by solving Eq.~\eqref{eq:bistability_threshold}. Lower values of $\gamma_m$ imply a more heterogeneous membership distribution. (a) The bistability threshold grows logarithmically or faster for $\gamma_m \leq 4$, but converges for $\gamma_m >4$. (b) The bistability threshold shows a maximum near $\gamma_m = 3$, suggesting that this is the maximum in the limit $m_\mathrm{max} \to \infty$.}
\label{fig:bistability_heterogeneous_membership}
\end{figure*}

In this section we investigate the effects of a heterogeneous membership distribution $g_m$ while keeping $p_n = \delta_{n,n_0}$ homogeneous to disentangle the impact of the different types of heterogeneity.
A first remark we can make about the invasion threshold [Eq.~\eqref{eq:invasion_threshold}] is that it is coherent with heterogeneous pair-approximation frameworks \cite{mata2014heterogeneous} on random networks when only dyadic interactions are considered, i.e., when $p_n = \delta_{n,2}$. In this case, we can set $\nu = 1$ without loss of generality, thus recovering the standard SIS model. The associated threshold is
\begin{align*}
    \lambda_\mathrm{c}^{\mathrm{SIS}} = \frac{\avg{m}}{\avg{m(m-1)}} \;,
\end{align*}
where $g_m$ can now be interpreted as the standard degree distribution of graphs.
This threshold, although quite accurate for most structures, does not capture the hub reinfection mechanism~\cite{st-onge2018phase}, and thus could be inaccurate for graphs with hubs of a very large degree.

More generally, for group interactions ($p_n \neq \delta_{n,2}$) we can see that a larger average excess membership \mbox{$\langle m (m-1) \rangle / \langle m \rangle$} always leads to a smaller invasion threshold $\lambda_\mathrm{c}$, akin to the standard SIS model, but the relationship is now nonlinear.
To see this, let us rewrite Eq.~\eqref{eq:invasion_threshold} as
\begin{align}\label{eq:invasion_threshold_other_form}
    \frac{1}{\avg{n}} \left \langle \sum_{i > 0}  \frac{ \;n! }{(n-i-1)! \; i!} \prod_{j = 1}^{i} \beta(n,j) \right \rangle = \frac{\avg{m} }{\avg{m(m-1)}} \;.
\end{align}
Since $\beta(n,i)$ is a monotonically increasing function of $\lambda$ for all $n,i$, then the left-hand side of Eq.~\eqref{eq:invasion_threshold_other_form} is a monotonically increasing function (of $\lambda$) as well. Consequently, if the right-hand side decreases, $\lambda_\mathrm{c}$ must decrease as well.

Assessing the impact of membership heterogeneity on the bistability threshold $\nu_\mathrm{c}$ is more complicated. In fact, Eq.~\eqref{eq:bistability_threshold} explicitly depends on the first three moments of $g_m$, but it also depends on the first two moments implicitly through $\lambda_\mathrm{c}$, at which $F$ must be evaluated.

In order to build our intuition, let us assume that we are able to keep fixed the first two moments $\langle m \rangle$ and $\langle m^2 \rangle$ while increasing $\langle m^3 \rangle$. This means that $\lambda_\mathrm{c}$ would not change, hence the only dependence on $g_m$ would be explicit in Eq.~\eqref{eq:bistability_threshold}.
Since the term depending on $\langle m^3 \rangle$ is negative, increasing the third moment implies that $F$ must increase if we want to balance Eq.~\eqref{eq:bistability_threshold}. But since $\mathrm{d}^2 \mathcal{M}/ \mathrm{d} r^2|_{r \to 0}$ increases with $\nu$ [see Fig.~\ref{fig:map}], and thereby $F$ as well, we can conclude that increasing $\langle m^3 \rangle$ while keeping the first two moments fixed leads to an increase of the bistability threshold $\nu_\mathrm{c}$.
This is validated in Fig.~\ref{fig:thirdmoment}, where we consider two distributions $g_m$ sharing the same first two moments, but a different third moment [see Fig.~\ref{fig:thirdmoment}(a)].
By comparing the associated epidemic curves in Fig.~\ref{fig:thirdmoment}(b), it is evident how the larger third moment suppresses the emergence of a bistable regime.

A corollary of this argument is that for certain structures, it is impossible to have bistability.
To see this, let us consider a power-law membership distribution of the form $g_m \sim m^{-\gamma_m}$.
In this case, since the bistability threshold depends on the third moment of $g_m$, while the invasion threshold only depends on the first two, by setting the exponent $3 < \gamma_m < 4$, the invasion threshold converges to a value $\lambda_\mathrm{c} > 0$, but $\nu_\mathrm{c}$ does not exist. In other words, it is impossible to have a discontinuous phase transition.

This second statement is validated in \ref{fig:bistability_heterogeneous_membership}(a), where we show that that the bistability threshold $\nu_\mathrm{c}$ appears to grow without bound as $m_\mathrm{max} \to \infty$ for $\gamma_m \leq 4$.
Instead, for $\gamma_m \geq 4$, the bistability threshold appears to converge, as expected, since the first three moments of $g_m$ converge as well.
What is more surprising is the non-monotonic behavior of $\nu_\mathrm{c}$ with respect to $\gamma_m$, which we  present in Fig.~\ref{fig:bistability_heterogeneous_membership}(b). The bistability threshold has a well defined maximum at a value of $\gamma_m$ that appears to converge to $\gamma_m = 3$ for $m_\mathrm{max} \to \infty$.
In other words, $\gamma_m = 3$ is the optimal value of membership exponent in suppressing the emergence of a discontinuous phase transition and the related bistability.

This can be understood from Eq.~\eqref{eq:bistability_threshold}: for $\gamma_m > 3$, the invasion threshold does not vary much since the first two moments of $g_m$ are finite. Hence maximizing the third moment maximizes $\nu_\mathrm{c}$, which corresponds to $\gamma_m \to 3$.
One could still be surprised that the bistability threshold grows more slowly with $m_\mathrm{max}$ in the range $2 < \gamma_m < 3$, since the invasion threshold $\lambda_\mathrm{c}$ tends toward zero.
In this case, the bistable regime exists, but its width $(\lambda_\mathrm{p},\lambda_\mathrm{c})$ simply vanishes as $\lambda_\mathrm{c} \to 0$.

\subsubsection{Heterogeneous group sizes}

\begin{figure*}[thb]
\begin{center}
    \includegraphics[width = \linewidth ]{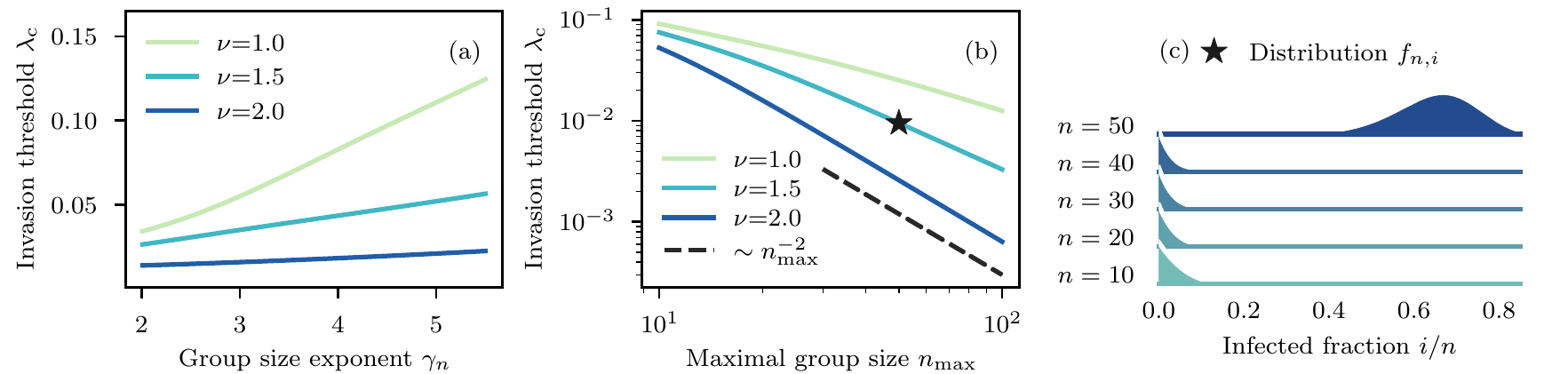}
\end{center}
\caption{{\bf Impact of heterogeneous group sizes on the invasion threshold.} We considered hypergraphs with power-law group size distributions $p_n \sim n^{-\gamma_n}$ with various exponents $\gamma_n$, a regular membership distribution $g_m = \delta_{m,4}$, and various exponents $\nu$ for the infection function in Eq.~\eqref{eq:power_law_infection}. We obtain the invasion threshold using Eq.~\eqref{eq:invasion_threshold}. Lower values of $\gamma_n$ implies a more heterogeneous group size distribution. (a) For a fixed $n_\mathrm{max} = 20$, the invasion threshold increases with larger $\gamma_n$, but the effect is more limited for larger $\nu$. (b) For a fixed $\gamma_n = 3$, the invasion threshold decreases like $n_\mathrm{max}^{-\nu}$ for large $n_\mathrm{max}$, indicating the onset of mesoscopic localization \cite{st-onge2021social,st-onge2021master}. (c) Stationary distribution $f_{n,i}$ for the starred case in (b), i.e., $\gamma_n = 3$ and $\nu = 1.5$, with $\lambda = 1.1\lambda_\mathrm{c}$, illustrating localization in the largest groups.}
\label{fig:invasion_heterogeneous_groups}
\end{figure*}

\begin{figure*}[thb]
\begin{center}
    \includegraphics[width = \linewidth ]{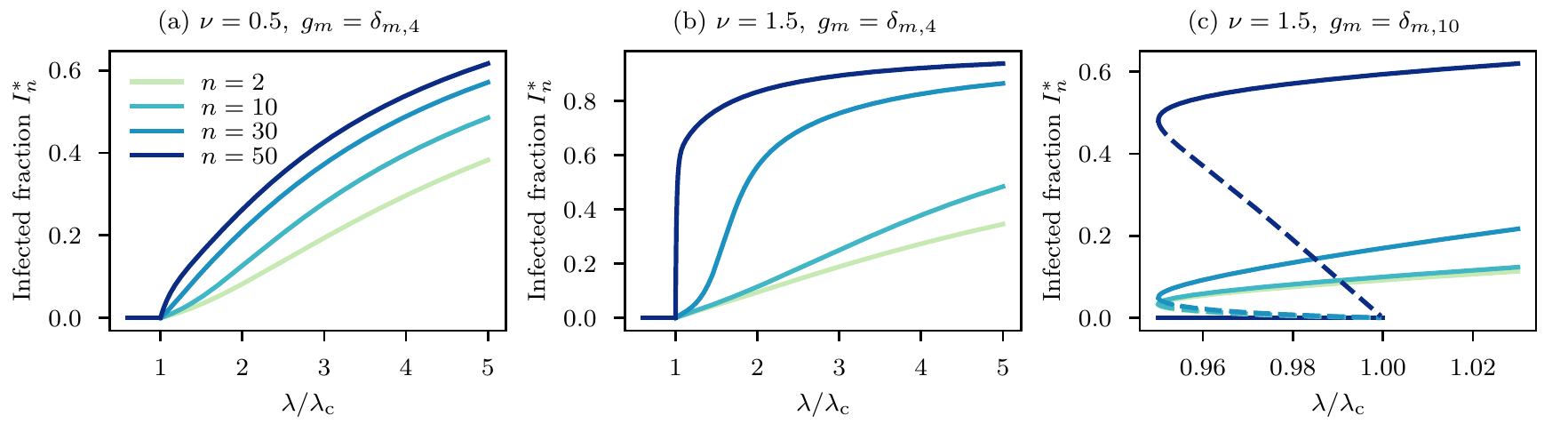}
\end{center}
\caption{{\bf Mesoscopic localization in large groups, illustrated by $\boldsymbol{I_n^*}$, the average stationary fraction of infected nodes in groups of various size $\boldsymbol{n}$.} We considered hypergraphs with power-law group size distributions $p_n \sim n^{-\gamma_n}$ with $\gamma_n = 3$, regular membership distributions of the form $g_m = \delta_{m,m_0}$, and various exponents $\nu$ for the infection function in Eq.~\eqref{eq:power_law_infection}. Near the invasion threshold $\lambda_\mathrm{c}$, $I_n^*$ is larger for larger groups in (a) with $\nu = 0.5$, but localization is much more pronouced for (b) with $\nu = 1.5$. (c) For discontinuous phase transitions, mesoscopic localization is still possible, but near $\lambda_\mathrm{c}$ we must look at the unstable solution for $I_n^*$ in the bistable regime.}
\label{fig:localization_phase_diagram}
\end{figure*}

\begin{figure*}[thb]
\begin{center}
    \includegraphics[width = \linewidth ]{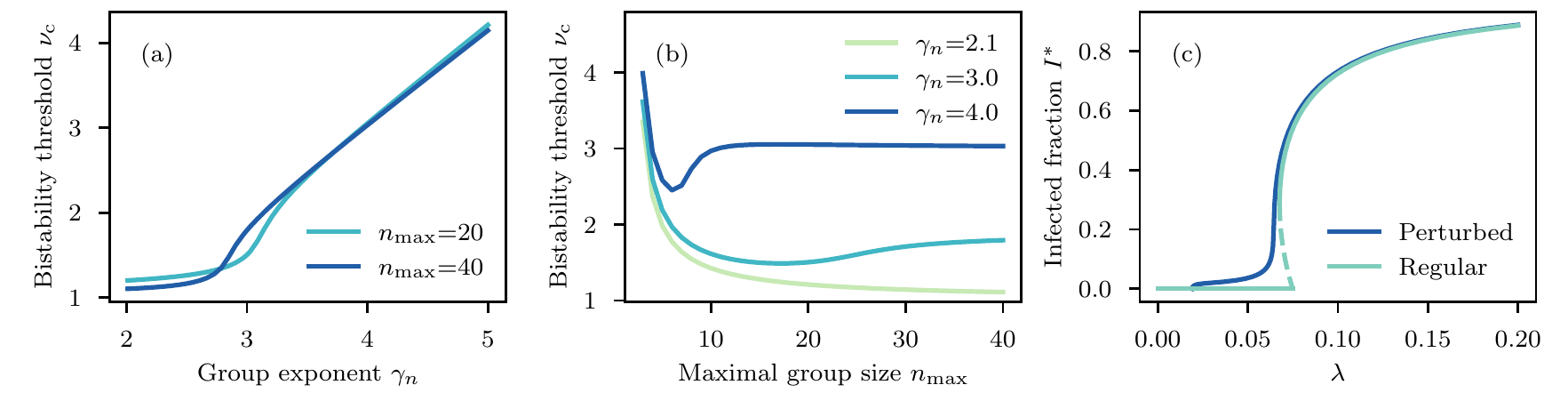}
\end{center}
\caption{{\bf Impact of heterogeneous group sizes on the bistability threshold.} We considered hypergraphs with power-law group size distributions $p_n \sim n^{-\gamma_n}$ with various exponents $\gamma_n$ and a regular membership distribution $g_m = \delta_{m,4}$. We solve the bistability threshold using Eq.~\eqref{eq:bistability_threshold}. Lower values of $\gamma_n$ implies a more heterogeneous group size distribution. (a) For a fixed $n_\mathrm{max}$, the bistability threshold increases with larger $\gamma_n$. (b) For a fixed $\gamma_n$, the bistability threshold has a non-monotonic relationship with $n_\mathrm{max}$. (c) Phase transition using $g_m = \delta_{m,4}$ and $\nu = 2.3$. We use $p_n = \delta_{n,4}$ for the regular case, and $p_n = (1-\epsilon) \delta_{n,4} + \epsilon \delta_{n,15}$ with $\epsilon = 10^{-3}$ for the perturbed case. Mesoscopic localization \cite{st-onge2021social,st-onge2021master} inhibits bistability in the perturbed case.}
\label{fig:bistability_heterogeneous_groups}
\end{figure*}

Let us now consider hypergraphs with heterogeneous group size distribution $p_n$, and homogeneous membership distribution, namely $g_m = \delta_{m,m_0}$.
In this case, the invasion threshold, as defined by Eq.~\eqref{eq:invasion_threshold}, depends on the whole distribution $p_n$, which makes drawing general conclusions on the impact of a heterogeneous distribution $p_n$ more difficult.

To get some intuitions, let us consider the standard SIS model, i.e., the case $\nu = 1$ in Eq.~\eqref{eq:power_law_infection}.
With our AMEs, it was shown that \cite{st-onge2021master}
\begin{align}\label{eq:invasion_threshold_sis}
    \lambda_\mathrm{c}^{-1} \simeq \left (\frac{\left \langle  m(m-1) \right \rangle }{\langle m \rangle }\right ) \left (\frac{\left \langle  n(n-1) \right \rangle}{\langle n \rangle} \right ) + n_\mathrm{max} \;,
\end{align}
for power-law distributions $p_n \sim n^{-\gamma_n}$ with large cut-offs $n_\mathrm{max}$.
The first term on the right-hand side of Eq.~\eqref{eq:invasion_threshold_sis} suggests that more heterogeneous groups-size distributions $p_n$ (smaller values of $\gamma_n$) lead to smaller invasion thresholds. Intuition tells us that we should expect this behavior for $\nu \neq 1$ as well.
We have therefore investigated numerically in Fig.~\ref{fig:invasion_heterogeneous_groups}(a) the invasion threshold as a function of the group size exponent for different values of $\nu$, confirming that more heterogeneous group sizes (smaller $\gamma_n$) do lead to a smaller invasion threshold, even for nonlinear infection functions ($\nu \neq 1$).
However, this effect is mitigated when larger values of $\nu$ are considered.
For large $\nu$ and large $n_\mathrm{max}$, the value of the invasion threshold is dominated by the cut-off, and scales as $\lambda_\mathrm{c} \sim n_\mathrm{max}^{-\nu}$, as illustrated in Fig.~\ref{fig:invasion_heterogeneous_groups}(b).

This behavior can be attributed to the onset of \textit{mesoscopic localization}. In Refs.~\cite{st-onge2021social,st-onge2021master}, it was shown analytically for $\nu = 1$ that, for certain combinations of ($\gamma_m,\gamma_n$), the epidemic near the invasion threshold [$\lambda = \lambda_\mathrm{c}(1 + \epsilon)$ with $\epsilon \ll 1$] is  dominated by the largest most influential groups. In these scenarios, the second term on the right-hand side in Eq.~\eqref{eq:invasion_threshold_sis} dominates the first one and, near $\lambda_\mathrm{c}$, the group prevalence $I_n$ grows exponentially with $n$, i.e., $I_{n_\mathrm{max}}/I_2 = \Omega\left(e^{a n_\mathrm{max}} \right)$ for some constant $a$.
While an analytical characterization of mesoscopic localization in the general case of $\nu \neq 1$ is out of the scope of this paper, we provide clear numerical evidence of localization phenomena in Fig.~\ref{fig:invasion_heterogeneous_groups}(c). The stationary distributions of the fraction of infected nodes in groups of increasing size $n$ is concentrated in the largest group ($n=50$) near the invasion threshold $\lambda_\mathrm{c}$.

Since mesoscopic localization was characterized using a linear contagion ($\nu = 1$) and a continuous phase transition \cite{st-onge2021social,st-onge2021master}, two natural questions arise: How does $\nu \neq 1$ affects localization? And what happens in the context of discontinuous phase transitions?
In Fig.~\ref{fig:localization_phase_diagram}, we present the phase diagram of the group prevalence $I_n$ for different scenarios. Comparing Fig.~\ref{fig:localization_phase_diagram}(a) and Fig.~\ref{fig:localization_phase_diagram}(b), we see that increasing $\nu$ from 0.5 to 1.5 (while keeping $g_m=\delta_{m,4}$) strengthens localization effects, which is expected since reinforcement effects are more important when the group prevalence is high.
In Fig.~\ref{fig:localization_phase_diagram}(c), we show a similar diagram, but for a discontinuous phase transition.
We see that the concentration of infected nodes in the largest groups is still possible, but the phenomenon is now associated with the \textit{unstable} solution near the invasion threshold [$\lambda = \lambda_\mathrm{c}(1 - \epsilon)$ with $\epsilon \ll 1$].
Therefore, mesoscopic localization affects both continuous and discontinuous phase transition with a bistable regime, but the exponential growth of $I_n$ with $n$ near $\lambda_\mathrm{c}$ concerns the stable solution in the former and the unstable solution in the latter.

If we now reinterpret the results of Fig.~\ref{fig:invasion_heterogeneous_groups} in light of these considerations, larger values of $\nu$ facilitate the onset of mesoscopic localization, where the largest groups drive the onset of the endemic phase, and make the invasion threshold scale as $\lambda_\mathrm{c} \sim n_\mathrm{max}^{-\nu}$. This explains why $\lambda_\mathrm{c}$ varies only slightly with $\gamma_n$ for $\nu = 2$ in Fig.~\ref{fig:invasion_heterogeneous_groups}(a).

In Fig.~\ref{fig:bistability_heterogeneous_groups}, we finally investigate the role of heterogeneous group sizes on the bistability threshold by varying the group exponent $\gamma_n$ and the maximal group size $n_\mathrm{max}$.
From Fig.~\ref{fig:bistability_heterogeneous_groups}(a), we see that a more heterogeneous group distribution, thereby increasing the fraction of larger groups, decreases the value of the bistability threshold $\nu_\mathrm{c}$. This is consistent with our observation on regular structures [Fig.~\ref{fig:bifurcation_hom}], for which larger groups appear to promote bistability. However, Fig.~\ref{fig:bistability_heterogeneous_groups}(b) brings some nuance to this statement: for a fixed exponent $\gamma_n$, there is an interesting non-monotonic relationship between $\nu_\mathrm{c}$ and the largest group $n_\mathrm{max}$. As such, the presence of larger groups does not always promote bistability.

We can again attribute this behavior to localization effects. In fact, we are able to illustrate this via a very simple example in Fig.~\ref{fig:bistability_heterogeneous_groups}(c). We look at the phase transition for a regular hypergraph with a fixed group size, $p_n = \delta_{n,4}$, and a perturbed version of it, where we introduce a small proportion of larger groups, $p_n = (1-\epsilon) \delta_{n,4} + \epsilon \delta_{n,15}$ with $\epsilon = 10^{-3}$.
For the regular distribution, the phase transition is discontinuous, while for the perturbed distribution it is continuous, with the contagion localized in the largest groups near the invasion threshold.
The bistability threshold $\nu_\mathrm{c}$ is larger for the perturbed distribution since mesoscopic localization reduces considerably the invasion threshold $\lambda_\mathrm{c}$.
The largest most influential groups drive and self-sustain an endemic state for smaller values of $\lambda$, hence preventing a bistable regime.

\subsection{Influence maximization}
\label{sec:influence_maximization}

Influence maximization broadly refers to the problem of selecting a subset of nodes to initially spark a diffusion process in order to maximize the effect.
The process could represent the spread of information, the diffusion of innovations or a viral marketing campaign \cite{rogers2010diffusion, domingos2001mining}.

There is a large body of literature on influence maximization in complex networks, where various models have been used: threshold models \cite{kempe2003maximizing,dodds2009threshold,morone2015influence,chen2010scalable}, independent cascade \cite{kempe2003maximizing}, and simple contagion models (SI, SIS, SIR) \cite{kitsak2010identification,chen2012identifying,erkol2019systematic,poux2020influential}, to name a few. Recently, these ideas have been also exported to higher-order networks \cite{amato2017influence, zhu2019influence}.

The effectiveness of an influence maximization procedure is often measured by the fraction of affected nodes (in the limit $t \to \infty$) for processes that terminate.
However, because the final epidemic size in the SIS dynamics does not depend on the seeds (other than for stochastic extinction), we will consider the simpler task of maximizing $\dot{I}(0)$, the initial spreading speed.
This is often a straightforward task to solve for graphs. Considering the SIR model for instance, one just needs to maximize the number of outgoing edges from infected to susceptible nodes, which implies that nodes of maximal degree would be optimal influencers.
However, we will show that additional considerations need to be accounted for in higher-order networks.
More specifically, our goal is to use our formalism to answer the following question: \textit{Should we focus on finding influential nodes, or seed the spread from influential groups?}

In this section, to simplify the notation, all dynamic quantities are evaluated at $t = 0$, e.g. $I(0) \equiv I$.

Let us assume that we are given a fixed hypergraph and an initial fraction of nodes that can be infected at the initial time $I = \epsilon \ll 1$ (the seeds of the contagion). Our task is to invade the system as fast as possible by maximizing $\dot{I}$ for a hypergraph contagion, which is equivalent to maximizing the objective function
\begin{align}
    \label{eq:inf_max_obj_func}
    \Phi[\mathcal{S},\mathcal{F}] = r \langle m s_m \rangle \;,
\end{align}
where we define the initial node states $\mathcal{S} \equiv \cbrace{s_m}_{m = 1}^{m_\mathrm{max}}$ and the initial group states $\mathcal{F} \equiv \cbrace{f_{n,i}| 0\leq  i \leq n}_{n = 2}^{n_\mathrm{max}}$. The optimization problem is also constrained by
\begin{subequations}
\label{eq:inf_max_const}
\begin{align}
     0 \leq s_m &\leq 1 \; \forall m \;, \\
     0 \leq f_{n,i} &\leq 1 \;\forall n,i \;, \\
    \sum_m s_m g_m &= 1-\epsilon \;,\label{eq:inf_max_const_3}\\
    \sum_{i}  f_{n,i}  &= 1 \; \forall n \;, \\
    \frac{\langle m s_m \rangle }{\avg{m}} &= \frac{1}{\avg{n}}\sum_{n,i} (n-i) f_{n,i} p_n \;. \label{eq:inf_max_const_5}
\end{align}
\end{subequations}
While the first four constraints come from the definitions of the variables, the last one is less straightforward. Equation~\eqref{eq:inf_max_const_5} ensures the consistency between $\mathcal{S}$ and $\mathcal{F}$, more specifically that the fraction of all memberships stubs belonging to susceptible nodes [left-hand side of Eq.~\eqref{eq:inf_max_const_5}] matches the fraction of susceptible nodes in groups [right-hand side of Eq.~\eqref{eq:inf_max_const_5}].

By combining the constraint of Eq.~\eqref{eq:inf_max_const_5} with the definition of $r$ as given by Eq.~\eqref{eq:mean-field1}, the objective function can be simplified as
\begin{align}
    \label{eq:inf_max_obj_func_2}
    \Phi[\mathcal{S},\mathcal{F}] \propto \sum_{n,i} \beta(n,i) (n-i) f_{n,i} p_n  \;.
\end{align}
Although it appears to be independent of $\mathcal{S}$, it depends on it implicitly through Eq.~\eqref{eq:inf_max_const_5}.

It is worth stressing that our formalism assumes that the membership stubs of nodes are assigned to groups uniformly at random, and thus we cannot \textit{engineer} both $\mathcal{S}$ and $\mathcal{F}$, i.e., choose at the same time the seeds according to their membership and the repartition of the seeds among the various group sizes.
Indeed, if we decide for instance to infect only nodes of a certain membership $m'$ \textit{and} we try to engineer $\mathcal{F}$, there are no guarantees we can achieve such configuration in practice---e.g. we cannot infect a node in a group if none of its nodes have membership $m'$.

We therefore compare two strategies to optimize the early spread:

\begin{enumerate}[A)]
    \item The \textit{influential spreaders} strategy: we engineer $\mathcal{S}$, i.e., we choose the fraction of seeds to assign to each membership class, and we assume a random configuration for the groups, i.e., all $\cbrace{f_{n,i}}_{i = 0}^n$ are binomial distributions with probability $q$ (to be determined).
    \item The \textit{influential groups} strategy: we engineer $\mathcal{F}$, i.e., we assign a certain number of seeds in the groups depending on their sizes, and assume that nodes are infected at random through the group to which they belong.
\end{enumerate}

\subsubsection{Influential spreaders}

In this strategy, we are free to engineer $\mathcal{S}$ in order to maximize $\Phi$, with respect to the constraints of Eq.~\eqref{eq:inf_max_const}.
Let us assume that $f_{n,i}$ is a binomial distribution,
\begin{align*}
    f_{n,i} = \binom{n}{i}q^i (1-q)^{n-i} \;.
\end{align*}
Using Eq.~\eqref{eq:inf_max_const_5}, we can identify
\begin{align*}
    q = 1 - \frac{\langle m s_m \rangle}{\langle  m \rangle} \;.
\end{align*}

An optimal solution $\mathcal{S}^\smallstar$ can be found by first finding the value $q^\smallstar$ that maximizes the objective function Eq.~\eqref{eq:inf_max_obj_func_2}, and then identifying any set $\mathcal{S}$ that satisfies the relation for $q = q^\smallstar$ above.

There are in general many optimal solutions possible, but they collapse into a single one when $q$ is sufficiently small, which is reasonable for $\epsilon \ll 1$.
In this case, we simply have that $\Phi \approx q$, and the optimal solution is intuitive: one needs to infect nodes of maximal membership first in order to maximize $q$.
Interestingly, this is true irrespective of $\beta(n,i)$, $p_n$ and $g_m$.

The infection function and the structure affect the maximal value of $\epsilon$ such that this solution is unique and optimal.
For example, in the simplest case of linear contagion, where $\beta(n,i) \propto i$, it is possible to show that this strategy is optimal up to $q = 1/2$ for all $g_m$ and $p_n$, and we expect even higher values for $\nu > 1$.
For all practical purposes, targeting nodes of highest membership is optimal, and this is the case in all experiments we considered (see Sec.~\ref{sec:experiments}).

\subsubsection{Influential groups}

In this second strategy, we want to engineer $\mathcal{F}$ in order to maximize $\Phi$ with respect to the constraints  Eq.~\eqref{eq:inf_max_const}.
Let us assume that we can do so by choosing a certain number of groups and infecting a certain portion of their nodes.
Following this procedure, one can realize that not all sets $\mathcal{F}$ satisfying Eq.~\eqref{eq:inf_max_const} are allowed.
For instance, if we decide to infect $i$ nodes in all groups of size $n$, the outcome is different from just having $f_{n,i} = 1$.
Indeed, nodes belong to more than one group, hence we need to account for spillover effects---groups of size $n' \neq n$ would have some infected nodes as well, and more than $i$ nodes could be infected in some groups of size $n$.

To do so, let us first define $\tilde{f}_{n,i}$ as the fraction of all the groups of size $n$ for which we infect $i$ nodes at random.
Note that if a node belongs to multiple groups, it can be \textit{chosen} more than once for infection, but the duplicates have no effect.
Spillovers are taken into account by considering that each of the $n-i$ nodes that have \textit{not} been chosen for infection in a group of size $n$ could have been infected in another group, with probability $u$ (to be determined).
Therefore, we can write
\begin{align}
    \label{eq:fni_from_tilde}
    f_{n,i} &= \sum_{j = 0}^i \tilde{f}_{n,i-j} B_{n-i+j,j} \;,
\end{align}
where
\begin{align*}
    B_{k,j} = \binom{k}{j} u^j (1-u)^{k-j} \;.
\end{align*}

Second, let us define $\eta$ as the fraction of all spots in groups that have been chosen for infection,
\begin{align}\label{eq:eta_def}
    \eta \equiv \frac{1}{\langle n \rangle} \sum_{n,i} i \tilde{f}_{n,i} p_n \;.
\end{align}
Since nodes within groups are chosen at random, a node of membership $m$ is susceptible if it has not been chosen for infection in any of the groups to which it belongs, i.e.,
\begin{align*}
    s_m = (1-\eta)^m \;.
\end{align*}
As a consequence, $\eta$ is constrained by Eq.~\eqref{eq:inf_max_const_3},
\begin{align*}
    \left \langle (1-\eta)^m \right \rangle  = 1 - \epsilon \;.
\end{align*}

The probability $u$ still needs to be obtained. It corresponds to the fraction of all memberships that are not matched with a spot chosen for infection in a group, but that are still associated with an infected node:
\begin{align*}
    u = \frac{\langle m (1 - s_m) \rangle - \eta \langle m \rangle}{(1- \eta) \langle m \rangle} \;.
\end{align*}

With this formulation, we engineer $\mathcal{F}$ indirectly through $\tilde{\mathcal{F}} = \lbrace \tilde{f}_{n,i} | 0 \leq i \leq n \rbrace_{n = 2}^{n_\mathrm{max}}$.
The objective function can be rewritten
\begin{align*}
    \Phi &\propto \sum_{n,i} \sum_{j = 0}^{i} \beta(n,i) (n-i) \tilde{f}_{n,i-j} B_{n-i+j,j} p_n  \;,\\
         &= \sum_{n,i} \sum_{j = 0}^{n-i} \beta(n,i+j) (n-i-j) \tilde{f}_{n,i} B_{n-i,j} p_n  \;.
\end{align*}
Since the objective function is a linear function of each $\tilde{f}_{n,i}$, the optimization problem can be solved using linear programming.

However, there is an intuitive and more efficient way to solve this problem exactly. We just need to identify the most \textit{cost-effective} $\tilde{f}_{n,i}$ by looking at the effect on $\Phi$ of increasing $\tilde{f}_{n,i}$,
\begin{align*}
    E \equiv \pd{\Phi}{\tilde{f}_{n,i}}  \propto \sum_{j = 0}^{n-i}\beta(n,i+j) (n-i-j) B_{n-i,j} p_n \;,
\end{align*}
versus the \textit{cost} of increasing $\tilde{f}_{n,i}$, i.e., the variation of $\eta$
\begin{align*}
    C \equiv \pd{\eta}{\tilde{f}_{n,i}} \propto i p_n \;.
\end{align*}
The most cost-effective $\tilde{f}_{n,i}$ maximizes the ratio
\begin{align}
    \label{eq:cost-effective-ratio}
    R(n,i) = \frac{E}{C} =  \frac{1}{i} \sum_{j = 0}^{n-i}\beta(n,i+j) (n-i-j) B_{n-i,j} \;.
\end{align}
Obviously, $i = 0$ is always the most cost-effective for all $n$ (since it has zero cost), but to satisfy Eq.~\eqref{eq:eta_def}, we must also fill some $\tilde{f}_{n,i}$ with $i > 0$.

Optimal solutions tend to fill the $\tilde{f}_{n,i}$ with $i > 0$ that maximizes $R(n,i)$, especially for sufficiently small $\epsilon$.
A general solution can be obtained using algorithm \ref{alg:fni_opt} presented in Methods Sec.~\ref{app:influential_groups}, building on this idea of cost effectiveness.
In the worst case, the computational complexity to obtain an optimal solution $\mathcal{F}^\smallstar$ under the influential groups strategy is $\mathcal{O}(m_\mathrm{max} + n_\mathrm{max}^3)$ when using the method presented in Methods Sec.~\ref{app:influential_groups}, which is much more efficient than using a general-purpose linear-programming method.

Equation \eqref{eq:cost-effective-ratio} also gives us an intuition of what defines influential groups when trying to maximize the early spread. If $\epsilon \ll 1$, then $u \ll 1$, hence we have
\begin{align*}
    R(n,i) \approx \frac{\beta(n,i)(n-i)}{i} \;, \\
           \propto i^{\nu-1}(n-i) \;,
\end{align*}
when considering $\beta(n,i) = \lambda i^\nu$.
For simple contagions ($\nu = 1$), picking the largest group with a single seed \mbox{($i = 1$)} is always optimal.
For hypergraph contagions with $\nu > 1$, the largest groups are the most influential as well, but the optimal number of seeds is generally $i > 1$.
Hence, beyond its size, the initial configuration of a group determines whether or not it is influential.

\subsubsection{Experiments}
\label{sec:experiments}

\begin{figure*}[tb]
\begin{center}
    \includegraphics[width = \linewidth ]{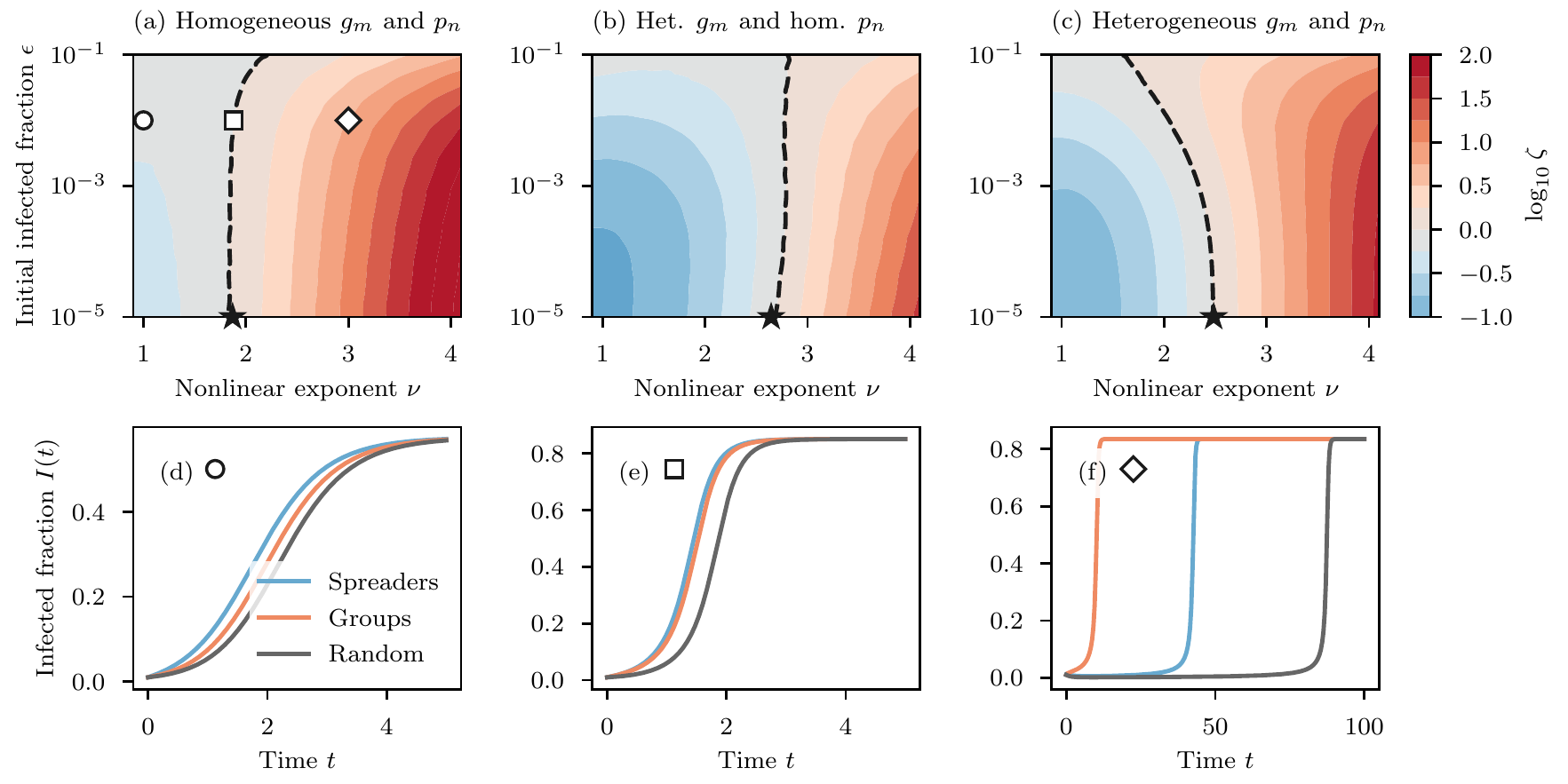}
\end{center}
\caption{{\bf Comparison of the influential spreaders and influential groups strategies.} (a)-(c) If $\log_{10} \zeta > 0$, this indicates that the influential groups strategy is better to maximize $\dot{I}$, and vice versa if $\log_{10} \zeta < 0$. We use different combinations of homogeneous and heterogeneous distributions. More specifically, for homogeneous distributions we use $g_m \propto a^me^{-a}/m!$ and $p_n \propto a^n e^{-a}/n!$, with $a = 5$, $m_\mathrm{min} = 1$, $n_\mathrm{min} = 2$, and $m_\mathrm{max} = n_\mathrm{max} = 20$. For heterogeneous distributions we use $g_m \propto m^{-\gamma_m}$ and $p_n \propto n^{\gamma_n}$, with $\gamma_m = \gamma_n = 3$, $m_\mathrm{min} = n_\mathrm{min} = 2$, and $m_\mathrm{max} = n_\mathrm{max} = 100$. The star markers indicate when $\zeta = 1$ in the limit $\epsilon \to 0$, using Eq.~\eqref{eq:zeta_lim_simpl}. Irregularities of the level curves are due to the discrete nature of $g_m$ and $p_n$, not to numerical errors. (d)-(f) Time evolution of the fraction of infected nodes for different strategies, with $\epsilon = 10^{-2}$ and $\nu \in \lbrace 1,1.88,3 \rbrace$, corresponding to the three empty markers in (a). We use $\lambda = 3\lambda_\mathrm{c}$ in each case. The random strategy corresponds to $s_m = 1 - \epsilon$ for all $m$, and $f_{n,i}$ is a binomial distribution with probability $\epsilon$.
}
\label{fig:inf_max_strat_comp}
\end{figure*}

To compare the influential spreaders and the influential groups strategies, we measure the ratio
\begin{align}
    \zeta \equiv \frac{\Phi_\mathcal{F}^\smallstar}{\Phi_\mathcal{S}^\smallstar} \;,
\end{align}
where $\Phi_\mathcal{F}^\smallstar$ and $\Phi_\mathcal{S}^\smallstar$ are the values of the objective function for the optimal solution of the influential groups and influential spreader strategies, respectively.
Therefore, $\zeta > 1$ indicates that the influential groups strategy is better to maximize $\dot{I}$, and vice versa if $\zeta < 1$.

In the Supplementary Information, we show that
\begin{align}
    \label{eq:zeta_lim}
    \lim_{\epsilon \to 0} \zeta = \frac{\beta(n',i')(n'-i') \langle n \rangle}{i' \langle \beta(n,1) n (n-1) \rangle m_\mathrm{max}} + \frac{\langle m(m-1) \rangle}{\langle m \rangle m_\mathrm{max}} \;,
\end{align}
where $(n',i')$ is the pair that maximizes the ratio $R(n,i)$, restricted to $i > 0$, in the limit $\epsilon \to 0$.
For general $\epsilon$, we need to solve numerically the optimization problem as discussed in the previous sections.

Interestingly, with $\beta(n,i) = \lambda i^\nu$, $\zeta$ is independent of $\lambda$, since $\Phi \propto \lambda$.
As a consequence, $\zeta$ is agnostic to the underlying phase of the system (healthy, bistable, or endemic).
Equation \eqref{eq:zeta_lim} simplifies to
\begin{align}
    \label{eq:zeta_lim_simpl}
    \lim_{\epsilon \to 0} \zeta = \frac{i'^{\nu-1}(n'-i') \langle n \rangle}{ \langle n (n-1) \rangle m_\mathrm{max}} + \frac{\langle m(m-1) \rangle}{\langle m \rangle m_\mathrm{max}} \;.
\end{align}

In Fig.~\ref{fig:inf_max_strat_comp}, we illustrate how $\zeta$ varies as we change $\nu$, $\epsilon$, and the underlying structure.
For homogeneous memberships and group sizes [Fig.~\ref{fig:inf_max_strat_comp}(a)], we see that the influential groups strategy performs better as soon as the contagion process is sufficiently nonlinear $(\nu \approx 2)$; for highly nonlinear contagions $(\nu \approx 4)$, the influential group strategy is much more effective, with $\zeta$ up to $100$.
When considering heterogeneous memberships, but still homogeneous group sizes [Fig.~\ref{fig:inf_max_strat_comp}(b)], the influential spreaders strategy performs better for moderately nonlinear contagions $(\nu \lesssim 2.8)$, otherwise the influential groups strategy is still a better choice.
Finally, considering a heterogeneous $p_n$ as well [Fig.~\ref{fig:inf_max_strat_comp}(c)] helps the performance of the influential groups strategy, especially for larger $\epsilon$.

When picking a pair $(\epsilon,\nu)$ far from $\zeta = 1$ [dashed lines in Figs.~\ref{fig:inf_max_strat_comp}(a), \ref{fig:inf_max_strat_comp}(b) and \ref{fig:inf_max_strat_comp}(c)], the strategy that maximizes $\dot{I}$ invades the system faster, as can be seen in Figs~\ref{fig:inf_max_strat_comp}(d)~and~\ref{fig:inf_max_strat_comp}(f). However, sufficiently close to $\zeta = 1$, maximizing $\dot{I}$ does not necessarily imply that $I(t)$ will be larger for all $t > 0$.
For instance, in Fig.~\ref{fig:inf_max_strat_comp}(e), $\zeta \approx 1$, but the influential spreader strategy is slightly better.
Therefore, one must be careful when interpreting the results of Fig.~\ref{fig:inf_max_strat_comp}.
One way to improve on our approach would be to consider higher-order temporal derivatives of $I$ to assess which strategy performs best, or refine the optimization procedure by trying to maximize these higher-order derivatives as well.

Interestingly, Figs.~\ref{fig:inf_max_strat_comp}(d)--(f) suggest that the initial speed, $\dot{I}(0)$, roughly correlates with the time taken by the disease to infect a given fraction of the population, a metric that has been used to measure influence for SI and SIR dynamics~\cite{karsai2011small, starnini2013immunization, lawyer2015understanding}.

Figure \ref{fig:inf_max_strat_comp}(f) also illustrates a particular feature of highly nonlinear contagions: the time to reach the stationary state can be excessively long for suboptimal strategies, despite $\lambda = 3 \lambda_\mathrm{c}$.
In this regime, the initial conditions have a much more important impact on the capacity of the contagion to invade the system, especially considering the possibility of stochastic extinction in real systems due to finite size.

\begin{figure}[tb]
\begin{center}
    \includegraphics[width = \linewidth ]{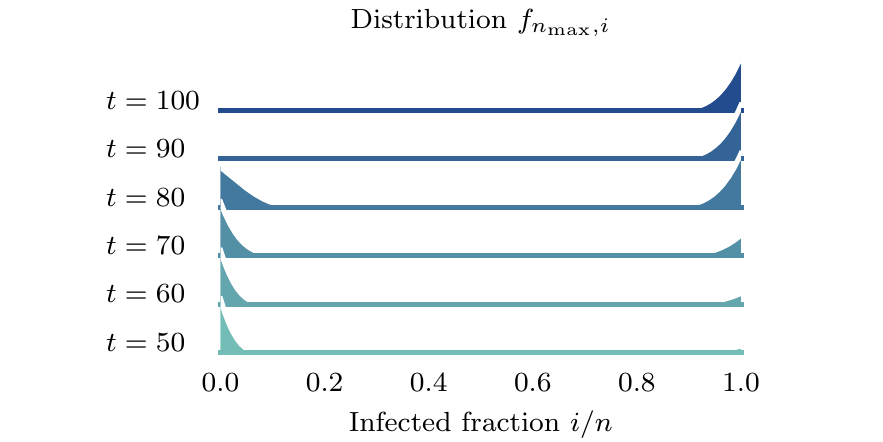}
\end{center}
\caption{{\bf Time evolution of the distribution $\boldsymbol{f_{n_\mathrm{max},i}}$ when using the random strategy.} The structure, the dynamics and the parameters are identical to the ones of Fig.~\ref{fig:inf_max_strat_comp}(f).}
\label{fig:time_evo_random}
\end{figure}

To understand why this is the case, we show in Fig.~\ref{fig:time_evo_random} the time evolution of $f_{n_\mathrm{max},i}(t)$, the distribution for the number of infected in the largest groups, when the nodes are initially infected at random.
For up to $t = 50$, only a few nodes are typically infected in the largest groups.
Then we see the apparition of a bimodal distribution around $t = 80$, where either almost all nodes of the largest groups are infected or almost all nodes are susceptible.
So instead of having a homogeneous transition, where all the largest groups have a similar state for all $t$, they diffuse asynchronously to states where almost all nodes are infected, at which point equilibrium is reached.
This transition is driven by stochastic fluctuations, which explains the slow take-off.
Conversely, the influential group strategy is by design optimizing the group configurations to facilitate this transition, which explains its good performance in Fig.~\ref{fig:inf_max_strat_comp}(f).

It is also worth mentioning that the transition in Fig.~\ref{fig:time_evo_random} hinges on the nonlinearity of the contagion process---as soon as the number of infected nodes is sufficiently large in a group, nonlinear effects boost infections, resulting in almost all nodes being infected.
Thus, nonlinear effects can be important \textit{locally}, even though the \textit{global} prevalence is still pretty low.
With mean-field approaches that ignore dynamical correlations, nonlinear effects---or contributions from higher-order interactions---kick in only for a sufficiently large global prevalence~\cite{li2021contagion}.

These results again highlight the importance of considering an accurate description of the inner dynamics of groups when studying hypergraph contagions.
In the context of influence maximization, optimizing group configurations is a crucial component; one should not focus exclusively on identifying the most central nodes.
Ultimately, an optimal strategy would capitalize on the synergy of these two important aspects.

\section{Discussion}

We have introduced group-based AMEs to describe hypergraph contagions.
Our  framework is analytically tractable, allowing us to obtain closed form implicit expressions for the critical and tricritical points. In addition, we have shown that it describes the dynamical process with remarkable accuracy when compared with Monte Carlo simulations.
Our formulation in terms of an infection rate function $\beta(n,i)$ makes it extremely flexible, allowing us to consider arbitrary group distribution with large group interactions, contrarily to existing HMF theories \cite{iacopini2019simplicial,landry2020effect,jhun2019simplicial} which instead require specifying  the rule for each different type of interaction separately.

Motivated by simplicity and recent results~\cite{st-onge2021bursty}, we analyzed in depth the consequences of a nonlinear infection rate function $\beta(n,i) = \lambda i^\nu$, highlighting the important role of influential groups in hypergraph contagions.

With our analytical results about the invasion and bistability thresholds, we were able to perform an exhaustive analysis of the phase transition and better understand the influence of a heterogeneous structure, both in terms of membership $m$ and group size $n$.
We found that the third moment of the membership distribution $g_m$ plays a crucial role, with large $\left \langle m^3 \right \rangle$ suppressing the onset of a discontinuous phase transition with a bistable regime, in line with other approaches~\cite{jhun2019simplicial,landry2020effect}.
This is best exemplified for power-law membership distributions $g_m \sim m^{-\gamma_m}$, where $\gamma_m = 3$ most suppresses bistability, and in the limit $m_\mathrm{max} \to \infty$, a discontinuous phase transition is only possible for $\gamma_m > 4$.

The phenomenon of mesoscopic localization~\cite{st-onge2021social,st-onge2021master}, driven by the most influential groups, also has important consequences on the phase diagram, with the effects being enhanced by superlinear infection $(\nu > 1)$.
In this case, the invasion threshold scales as $\lambda_\mathrm{c} \sim n_\mathrm{max}^{-\nu}$, and for $\lambda$ close to $\lambda_\mathrm{c}$, infected nodes are found almost exclusively in the largest groups.
This localization of the contagion thereby inhibits bistability by enforcing an endemic state with a very small global fraction of infected nodes.

Our approach, furthermore, provided novel insights concerning the problem of influence maximization for hypergraph contagions.
We focused on the problem of maximizing the early spread, and proposed two strategies: allocating seeds to the influential spreaders (engineering $s_m$), or to the influential groups (engineering $f_{n,i}$).
For various types of structures, the latter strategy performs better for contagions that are sufficiently nonlinear, highlighting the key role of influential groups on the transient state of the system.

For the process we considered, the notion of influential groups to seed and sustain hypergraph contagions are mostly aligned---in both cases, the largest groups typically have a dominant role.
In the case of influence maximization, however, we showed that a careful seed allocation is also essential to determine whether or not a group is influential.
Moreover, a more realistic infection function $\beta(n,i)$ that actually depends on $n$ could affect which groups are most influential in both scenarios.

Our work constitutes a first step towards a better understanding of the role of higher-order interactions on the outreach of information spreading~\cite{rogers2010diffusion}, and resonates with other recent theoretical findings on higher-order naming games, where big groups facilitate takeover of committed minorities in social convention~\cite{iacopini2021vanishing}.
Interestingly, AMEs thus provide an analytical avenue to study recent empirical results showing how social contagions and movements defy classic influence maximization.
As one example, networked counterpublics \cite{jackson2015hijacking} are public spaces used by underrepresented groups to gather legitimacy and form tight-knit communities.
Therein, non-dominant forms of knowledge can still spread and reach widespread attention through dense communities (influential groups) despite the limited connectivity of their members (non-influential spreaders).
These results provide one more addition to the mounting evidence that groups of elementary elements are the foundational unit of many complex systems.

Many avenues are now left open to explore and broaden the applicability of our group-based AMEs.
While we restrained ourselves to a particular nonlinear infection rate function $\beta(n,i)$ and a constant recovery rate, other dynamical processes could be considered, each having their own phenomenology and a rich dynamical behavior.
In the Supplementary Information for instance, we briefly discuss how our framework can be applied to threshold models of the form $\beta(n,i) = \delta_{n-1,i}$, but one could consider other traditional dynamical processes, such as voter models \cite{sood2005voter}.

In the Supplementary Information, we provide a roadmap to include \textit{structural} two-point correlations, but a thorough characterization of the impact of correlation patterns on bistability, mesoscopic localization and influence maximization is still lacking.
The inclusion of \textit{dynamical} correlation around nodes is a more tedious task that would require a fusion between degree-based~\cite{marceau2010adaptive,gleeson2011highaccuracy,gleeson2013binarystate} and group-based~\cite{hebert-dufresne2010propagation,osullivan2015mathematical,st-onge2021social,st-onge2021master} AMEs.
This would allow to describe almost exactly \textit{short-range} dynamical correlations, namely correlations between the states of nodes and their direct neighbors.
Incorporating \textit{long-range} correlations---beyond first neighbors---in AME frameworks, without a prohibitive computational time due to combinatorial explosion, is still an open problem.

Finally, many directions could be taken with regards to the influence maximization problem on hypergraphs.
One avenue would be to analyze the notion of influential groups and influential spreaders from the perspective of centrality measures for hypergraphs~\cite{tudisco2021node}.
Another would be to investigate the closely related problem of targeted immunization \cite{pastor-satorras2015epidemic,pastor-satorras2002immunization}.

\section{Methods}

\subsection{Contagion on real-world hypergraphs}\label{app:real-world}

\subsubsection{Simulation of contagions}\label{app:simulation}

We used a standard Gillespie algorithm for the simulation of contagions on hypergraphs.
We decompose the whole process into events $j \in J$, that each happens at rate $\omega_j$.
The next event to happen is chosen with probability
\begin{align*}
    P(j) = \frac{\omega_j}{\sum_{j \in J} \omega_j} \;,
\end{align*}
and the time step between two events $\Delta t$ is distributed exponentially with mean $\langle \Delta t \rangle = 1/\sum_{j \in J} \omega_j$.

There are two types of events: infection and recovery.
On the one hand, all susceptible nodes in a group can be considered equivalent with regards to infection. Consequently, each group is chosen for an infection event with rate
\begin{align*}
    \omega_\mathrm{inf}(n,i) = (n-i)\beta(n,i)\;.
\end{align*}
Once a group is chosen for an infection event, one of the $(n-i)$ susceptible nodes is chosen uniformly at random to become infected.
On the other hand, all infected nodes perform a recovery event with rate $\omega_\mathrm{rec} = 1$.

We store all possible events in an efficient data structure called a \texttt{SamplableSet} \cite{st-onge2018samplableset}, where insertion, deletion and sampling of elements (events) all have a computational complexity $\mathcal{O}\left [ \log \log \left ( \omega_\mathrm{max}/\omega_\mathrm{min} \right) \right]$ \cite{st-onge2019efficient}, where $\omega_\mathrm{max}$ and $\omega_\mathrm{min}$ are respectively the maximal and minimal rate among $\lbrace w_j \rbrace_{j \in J}$ .
This makes the sampling and the updating of the data structure extremely fast, which is especially useful when $\lbrace w_j \rbrace_{j \in J}$ spans multiple scales.

Once an event is performed---for instance a node recovers---we need to update the rate $\omega_\mathrm{inf}$ of all groups to which this node belongs.
This is the most costly part of the algorithm, which unfortunately cannot be overcome. This essentially means the simulation procedure is slower for hypergraphs with large average excess membership $\langle m(m-1) \rangle/\langle m \rangle$.

In Fig.~\ref{fig:simulation_sociopatterns} and \ref{fig:simulation_coauthorship}, we compare the stationary state solutions from our formalism with estimates from Monte Carlo simulations.
To compute estimates, we let the system relax during a burnin period $\tau_\mathrm{b} \in [10^2,10^4]$ then we sample $\mathcal{N} \in [10,10^4]$ states, both depending on the size of the hypergraph and if multiple randomized hypergraphs are being used. Sampled states are separated by a decorrelation period $\tau_\mathrm{d} = 1$.

To simulate contagions in the stationary state, we used two approaches, ordinary simulations and the quasistationary-state method.
\paragraph{Ordinary simulation method.}
With this approach, we simply let the simulation run and do not intervene. This is usually not the method of choice, especially for small hypergraphs near the invasion threshold, because finite size systems all eventually reach the absorbing state where all nodes are susceptible. This is, however, more practical to obtain the lower branch in Fig.~\ref{fig:simulation_sociopatterns}(f), or faster for large hypergraphs, as in Fig.~\ref{fig:simulation_coauthorship}.

\paragraph{Quasistationary-state method.}
This approach aims at sampling the quasistationary distribution of the contagion process \cite{deoliveira2005how}, which is defined as the probability distribution for all states in the limit $t \to \infty$, except for the absorbing state.
We used the state-of-the-art method discussed in Ref.~\cite{deoliveira2005how}, where we keep a history of past states (in our case up to 50 states).
We update the history by removing one uniformly at random and storing the current state after each decorrelation period $\tau_\mathrm{d} \in [0.1,1]$.
Each time the absorbing state is reached during the simulation, we pick a state from the history uniformly at random to replace the current one.
This method is well suited for finite-size analysis, and especially useful for simulations on small hypergraphs, such as in Fig.~\ref{fig:simulation_sociopatterns}.

\subsubsection{Datasets}\label{app:datasets}
The simulations shown in Fig.~\ref{fig:simulation_sociopatterns} and Fig.~\ref{fig:simulation_coauthorship} run on two different empirical social structures that encode different types of social higher-order interactions. Here, we briefly describe the nature of these two datasets and the techniques used to construct the associated higher-order structures.

\paragraph{Face-to-face interactions in a French primary school.}
Originally collected in Ref.~\cite{stehle2011highresolution} as part of the {\it SocioPatterns}~\cite{facetofacedata} collaboration, this dataset contains information of face-to-face interactions between children of a primary school in Lyon recorded over 2 days. Participants are given wearable sensors (placed on their chests), and a contact is detected whenever two sensors are in close proximity (1.5m). The initial temporal resolution of this dataset is 20s, but contacts have been further pre-processed as in Ref.~\cite{iacopini2019simplicial} in order to construct a static hypergraph from the temporal sequence of interactions. In particular, considering each child as a node, we aggregated different snapshots using a temporal window of 15 minutes and computed all the maximal cliques appearing in each window. Cliques were then aggregated across the entire time range, retaining only those that appeared at least 3 times, and finally ``promoted" to groups. Some properties of the obtained structure are reported in the caption of Fig.~\ref{fig:simulation_sociopatterns}.

\paragraph{Co-authorship in computer science.}
{\it DBLP}~\cite{coauthorshipdata} is an online bibliography containing information on major computer science journals and proceedings. This dataset, already pre-processed in Ref.~\cite{benson2018simplicial} (from the release 3, 2017), consists in a list of publications and respective authors that naturally calls for higher-order representations~\cite{patania2017shape}. In particular, each author corresponds to a node and any collaboration of $n$ (co-)authors in a single publication corresponds to a group of size $n$. We constructed a hypergraph by aggregating all the resulting groups together, but without considering single author publications (these have been removed in order to avoid disconnected nodes). In addition, the original dataset contained $1\,831\,127$ nodes and $2\,954\,518$ groups, which is too large to perform simulations on a personal computer in a reasonable time. Therefore, we obtained a sub-hypergraph by performing a breadth-first search, starting from a random group, then visiting all groups at a maximum distance of 2 when considering the one-mode projection of the original hypergraph on the groups. This ensures that the resulting sub-hypergraph is connected. Some properties of the obtained structure are reported in the caption of Fig.~\ref{fig:simulation_coauthorship}.

\subsubsection{Randomization and data augmentation}\label{app:data-augmentation}

In Fig.~\ref{fig:simulation_sociopatterns} and \ref{fig:simulation_coauthorship}, we make use of randomized versions of the original hypergraph. In Fig.~\ref{fig:simulation_sociopatterns}, we also use \textit{expanded} versions, where the size of the network is increased by a factor $x$. In all cases, we use the same procedure ($x = 1$ if the hypergraph is not expanded).

Let us first note $\boldsymbol{m} = [m_1,m_2,\dots]$ and $\boldsymbol{n} = [n_1,n_1,\dots]$ the membership sequence and the group size sequence of the original hypergraph, i.e., the list for the membership of each node and the list for the size of each group. From these sequences, we create two expanded sequences $\boldsymbol{m}'$ and $\boldsymbol{n}'$, which are formed of $x$ copies of $\boldsymbol{m}$ and $\boldsymbol{n}$ respectively.
This can be seen as the membership and group size sequences for a hypergraph that is $x$ times larger.

For each expanded sequence, we create a stub list. For instance, for the node $j$ of the expanded hypergraph, we include $m_j'$ copies of the label $j$ in the stub list for the nodes. Similarly, we include $n_\ell'$ copies of the label $\ell$ in the stub list for the groups. By definition, these two stub lists are of the same length, $M'$, which corresponds to the number of \textit{edges} in the bipartite representation of the hypergraph. We can thus shuffle them and match the entries of both lists, thereby assigning nodes to groups---or equivalently creating edges between nodes and groups in the bipartite representation of the hypergraph.

We then remove multi-edges (nodes assigned multiple times to a same group) by performing edge swaps~\cite{fosdick2018configuring}. We then perform $M'$ additional edge-swap attempts at random---picking two random edges, swapping the groups, and accepting the swap if it does not create multi-edges. This ensures the uniformity of the generation process (see the Supplementary Information). The resulting hypergraph is a randomized version of the original hypergraph, expanded by a factor $x$.

\subsection{Influential groups solutions}
\label{app:influential_groups}

An intuitive approach to solve the problem would be to sort all pairs $(n,i)$ in decreasing order of their $R(n,i)$ values (for $i > 0$), then fill $\tilde{f}_{n,i}$  up to 1 following this order, until $I = \epsilon$, or more directly until $\eta$ reaches the value prescribed by $\epsilon$.
However, this approach does not account for the fact that one may encounter multiple times a same $n$ value before the condition $I = \epsilon$ is reached.
For instance, let us assume $(n,i)$ is the next the pair with the highest value $R(n,i)$, but there exists a pair $(n,i')$ with $R(n,i') \geq R(n,i)$ and we have already assigned $\tilde{f}_{n,i'} = 1$. What is the best option?
\begin{enumerate}
    \item Discard the $(n,i)$ pair.
    \item Fill the associated $\tilde{f}_{n,i}$ up to 1 and decrease the value of $\tilde{f}_{n,i'}$ accordingly.
\end{enumerate}

It turns out that an optimal solution is constructed by choosing one or the other depending on certain conditions.
Option 1 is chosen whenever $i < i'$, because it can only reduce the total contribution to $\Phi$.
If $i > i'$, we assign a new cost-effective ratio to the pair $(n,i)$, accounting for the fact that we need to decrease $\tilde{f}_{n,i'}$:
\begin{align*}
    \hat{R}(n,i) &= \frac{iR(n,i) - i'R(n,i')}{i-i'} \;.
\end{align*}
This can be interpreted as the cost-effective ratio for the additional infected nodes $(i-i')$ that we add to the configuration.
Note that $\hat{R}(n,i)$ can be negative, which is not a problem: this only means that infecting these nodes decreases the objective function $\Phi$.
If $\hat{R}(n,i)$ is still the highest ratio when compared with the ratios from other available pairs, option 2 is chosen.

The procedure to find an optimal set $\tilde{\mathcal{F}}^\smallstar$ is presented in algorithm \ref{alg:fni_opt}, where we use a priority queue $Q$ to keep sorted the pairs $(n,i)$ in decreasing order of their cost-effective ratio.
Using a binary heap for $Q$, inserting any pair $(n,i)$ or removing the pair with maximum cost-effective ratio takes $\mathcal{O}(\log |Q|)$ elementary operations in the worst case, where $|Q|$ is the number of elements in the queue.
Since $|Q| \leq n_\mathrm{max}^2$, insertions and removals have a complexity $\mathcal{O}(\log n_\mathrm{max})$.
In the worst case, we will perform $\mathcal{O}(n_\mathrm{max}^2)$ insertions and removals before the condition $I = \epsilon$ is reached, hence the worst-case complexity for algorithm \ref{alg:fni_opt} is $\mathcal{O}(n_\mathrm{max}^2 \log n_\mathrm{max})$.

We also need to precompute $R(n,i)$ for all $n,i > 0$ and get $\mathcal{F}^\smallstar$ from $\tilde{\mathcal{F}}^\smallstar$, both $\mathcal{O}(n_\mathrm{max}^3)$. Moreover, solving for $\eta$ is $\mathcal{O}(m_\mathrm{max})$, so the overall computational complexity of the influential groups optimization procedure is $\mathcal{O}(m_\mathrm{max} + n_\mathrm{max}^3)$.

\begin{algorithm}[H]
\caption{Optimization for influential groups}
\label{alg:fni_opt}
\textbf{Input:} $p_n$, $R(n,i)$, $\eta$ \\
\textbf{Output:} $\tilde{\mathcal{F}}^\smallstar$
\begin{algorithmic}[1]
    \State $\tilde{f}_{n,i} \gets 1$ if $i = 0$, else $0$, $\forall n,i$
    \State $\psi \gets \eta \langle n \rangle$ \Comment{Initial resources}
    \State $i_\mathrm{opt} \gets \emptyset$ \Comment{Associative array}
    \State $Q \gets \emptyset$ \Comment{Priority queue}
    \For {all $(n,i)$ s.t $n \leq n_\mathrm{max}$, $p_n > 0$ and $1 \leq i \leq n$}
        \State $R \gets R(n,i)$
        \State $Q \gets (n,i)$ with priority $R$
    \EndFor
    \While {$Q$ is not empty and $\psi > 0$}
        \State $(n,i) \gets Q$ \Comment{Highest priority element}
        \If{$n \notin i_\mathrm{opt}$}
            \State $\tilde{f}_{n,i} \gets \mathrm{min}[1, \psi/(ip_n)]$
            \State $\tilde{f}_{n,0} \gets \tilde{f}_{n,0} - \tilde{f}_{n,i}$
            \State $\psi \gets \psi - \tilde{f}_{n,i} i p_n$
            \State $i_\mathrm{opt} [n] \gets i$
        \ElsIf{$i > i_\mathrm{opt}[n]$}
            \State $i' = i_\mathrm{opt}[n]$
            \State $\hat{R} \gets [iR(n,i) - i'R(n,i')]/(i-i')$
            \If{$\hat{R} > R$ for all $R \in Q$}
                \State $\tilde{f}_{n,i} \gets \mathrm{min}\{1, \psi/[p_n(i-i')]\}$
                \State $\tilde{f}_{n,i'} \gets \tilde{f}_{n,i'} - \tilde{f}_{n,i}$
                \State $\psi \gets \psi - \tilde{f}_{n,i} p_n (i-i')$
                \State $i_\mathrm{opt}[n] \gets i$
            \Else
                \State $Q \gets (n,i)$ with priority $\hat{R}$
            \EndIf
        \EndIf
    \EndWhile
\end{algorithmic}
\end{algorithm}

\section*{Acknowledgments}
The authors acknowledge Calcul Qu\'{e}bec for computing facilities.
This work was supported by the Fonds de recherche du Qu\'{e}bec – Nature et technologies (G.S.), the Natural Sciences and Engineering Research Council of Canada (G.S., A.A.), the Sentinelle Nord program of Universit\'{e} Laval, funded by the Canada First Research Excellence Fund (G.S., A.A.), the United Kingdom Regions Digital Research Facility (RDRF)-Urban Dynamics Lab under EPSRC Grant No. EP/M023583/1 (I.I.), the James S. McDonnell Foundation $21^{\text{st}}$ Century Science Initiative Understanding Dynamic and Multi-scale Systems (I.I.), the Agence Nationale de la Recherche (ANR) project DATAREDUX [ANR-19-CE46-0008] (A.B. and I.I.), the National Science Foundation under grant No. OIA-2019470 (L.H.-D.), Google Open Source under the Open-Source Complex Ecosystems and Networks (OCEAN) project (L.H.-D.), and the National Institutes of Health 1P20 GM125498-01 Centers of Biomedical Research Excellence Award (L.H.-D.).

\subsection*{Data availability}
The datasets are available from the original sources:
\begin{itemize}
    \item Face to face interaction in a primary school~\cite{facetofacedata}.
    \item Co-authorship data in computer science~\cite{coauthorshipdata}.
\end{itemize}
Since the data have been pre-processed before being used in simulations, we also provide the pre-processed data with our code.

\subsection*{Code availability}
The code is available on Github~\cite{codesimplagion}.

\section*{Competing interests} The authors declare that they have no competing interests. The funders had no role in study design, data collection and analysis, decision to publish, or preparation of the manuscript.

\subsection*{Author contributions}

G.S., I.I., V.L., A.B., G.P., A.A., and L.H.-D. conceptualized the study. G.S. and L.H.-D. developed the model. G.S. derived the theoretical results, implemented the algorithms, and performed the numerical experiments. G.S., I.I., V.L., A.B., G.P., A.A., and L.H.-D. analyzed the results and contributed to writing the manuscript.


%

\clearpage


\section*{Supplementary material}

# Influential groups for seeding and sustaining nonlinear contagion in heterogeneous hypergraphs
## — Supplementary Information —

Guillaume St-Onge,[1, 2, *] Iacopo Iacopini,[3, 4, 5, 6] Vito Latora,[6, 7, 8, 9] Alain Barrat,[4, 10]
Giovanni Petri,[11, 12] Antoine Allard,[1, 2, 13] and Laurent Hébert-Dufresne[1, 13, 14, †]

[1]*Département de physique, de génie physique et d'optique,*
*Université Laval, Québec (Québec), Canada, G1V 0A6*
[2]*Centre interdisciplinaire en modélisation mathématique,*
*Université Laval, Québec (Québec), Canada, G1V 0A6*
[3]*Department of Network and Data Science, Central European University, 1100 Vienna, Austria*
[4]*Aix Marseille Univ, Université de Toulon, CNRS, CPT, Marseille, France*
[5]*Centre for Advanced Spatial Analysis, University College London, London, W1T 4TJ, United Kingdom*
[6]*School of Mathematical Sciences, Queen Mary University of London. Mile End Road, London E1 4NS, United Kingdom*
[7]*The Alan Turing Institute, The British Library, London NW1 2DB, United Kingdom*
[8]*Dipartimento di Fisica ed Astronomia, Università di Catania and INFN, I-95123 Catania, Italy*
[9]*Complexity Science Hub, Josefst ädter Strasse 39, A 1080 Vienna, Austria*
[10]*Tokyo Tech World Research Hub Initiative (WRHI), Tokyo Institute of Technology, Tokyo, Japan*
[11]*Mathematics and Complex Systems Research Area, ISI Foundation, Turin, Italy*
[12]*ISI Global Science Foundation, New York, USA*
[13]*Vermont Complex Systems Center, University of Vermont, Burlington, VT 05401, USA*
[14]*Department of Computer Science, University of Vermont, Burlington, VT 05401, USA*
(Dated: October 4, 2021)

## CONTENTS

## I. AME CORRECTION FOR CORRELATIONS

It is straightforward to adapt our formalism to include two-point structural correlations between node memberships and group sizes. Let $P(m, n)$ be the joint probability distribution for the membership of a node and the size of a group at the endpoint of an edge in the bipartite representation of the hypergraph. Marginalization results in

$$\sum_m P(m, n) = \frac{n p_n}{\langle n \rangle}$$

and

$$\sum_n P(m, n) = \frac{m g_m}{\langle m \rangle} \ .$$

---

[*] guillaume.st-onge.4@ulaval.ca
[†] Laurent.Hebert-Dufresne@uvm.edu

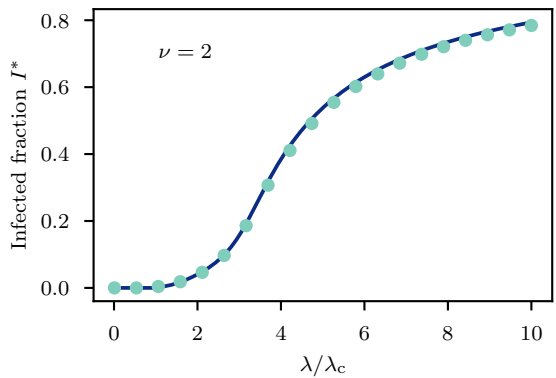

Supplementary Figure 1. **Assessing dynamical correlation around nodes by removing nodes with a large membership.** We performed the same experiment as in Fig. 3(c), but we used a homogeneous membership distribution $g_m \approx 0.25\delta_{m,3} + 0.75\delta_{m,4}$ with the same mean $\langle m \rangle \approx 3.75$. The group size distribution is unchanged.

We thus define $P(m|n)$, the conditional probability for a node to be of membership $m$ in a group of size $n$, and $P(n|m)$, the conditional probability for a group to be of size $n$ taking a random group to which a node of membership $m$ belongs,

$$P(m|n) \equiv \frac{P(m,n)\langle n \rangle}{np_n} \quad \text{and} \quad P(n|m) \equiv \frac{P(m,n)\langle m \rangle}{mg_m} \ . \tag{S1}$$

From these two definitions, it is sufficient to replace $r(t) \mapsto r_m(t)$ and $\rho(t) \mapsto \rho_n(t)$ into Eq. (4) to account for two-point structural correlations, where

$$r_m(t) = \frac{\sum_{n,i}\beta(n,i)n^{-1}(n-i)f_{n,i}P(n|m)}{\sum_{n,i}n^{-1}(n-i)f_{n,i}P(n|m)} \ , \tag{S2a}$$

$$\rho_n(t) = \frac{\sum_m (m-1)r_m s_m P(m|n)}{\sum_m s_m P(m|n)} \ . \tag{S2b}$$

While this correction leaves unaltered the dimension of the system of ordinary differential equations, the computational complexity associated to the numerical integration of Eq. (4) increases from $\mathcal{O}\left(n_{\max}^2 + m_{\max}\right)$ to $\mathcal{O}\left(n_{\max}^2 m_{\max}\right)$ per time step.

For a more in-depth study of the effect of structural correlations on hypergraph contagion models, see Ref. [1]. To include dynamical correlations around nodes, it would necessitate a much more thorough reformulation of our formalism, for instance by combining the partitioning in groups with the node partitioning typically used in degree-based AMEs [2–5].

## II. FURTHER ASSESSMENT OF THE EFFECT OF DYNAMICAL CORRELATION AROUND NODES

We conjecture that the disparity between the AMEs and the simulations in Fig. 3(c) is caused by the presence of nodes with very large membership. The dynamical correlations around these nodes become important at high prevalence and it is not correctly captured by our AMEs.

To support our claim, we reproduced the simulation and the results of the AMEs in Supplementary Figure 1, keeping the same group size distribution, but using a homogeneous membership distribution. The mean membership $\langle m \rangle \approx 3.75$ is preserved.

The phase diagram is quite different without the heterogeneous membership distribution—the phase transition is smeared due to mesoscopic localization near the invasion threshold. However, the point is that the agreement is better at high prevalence, which suggests that the nodes with large membership had a role to play in the disparity.

## III. JACOBIAN OF THE MEAN-FIELD MAP

All quantities are assumed to have reached the stationary sate. As shown in Sec. II B, the stationary state solutions are completely characterized by the mean field $r$. Therefore, one just needs to obtain solutions $r$ that respect

$$r = \mathcal{M}[\rho(r)] \ .$$

A useful quantity to find the fixed points numerically and analyze their stability is the Jacobian

$$J_r = \frac{\mathrm{d}\mathcal{M}}{\mathrm{d}r} = \frac{\mathrm{d}\mathcal{M}}{\mathrm{d}\rho}\frac{\mathrm{d}\rho}{\mathrm{d}r} \ .$$

For the first term, we define the following quantities

$$u_1(\rho) = \sum_{n,i} \beta(n,i)(n-i)f_{n,i}(\rho)p_n \ ,$$

$$v_1(\rho) = \sum_{n,i} (n-i)f_{n,i}(\rho)p_n \ ,$$

$$\frac{\mathrm{d}u_1}{\mathrm{d}\rho} = \sum_{n,i} \beta(n,i)(n-i)\frac{\mathrm{d}f_{n,i}}{\mathrm{d}\rho}p_n \ ,$$

$$\frac{\mathrm{d}v_1}{\mathrm{d}\rho} = \sum_{n,i} (n-i)\frac{\mathrm{d}f_{n,i}}{\mathrm{d}\rho}p_n \ ,$$

and write

$$\frac{\mathrm{d}\mathcal{M}}{\mathrm{d}\rho} = \frac{1}{v_1}\frac{\mathrm{d}u_1}{\mathrm{d}\rho} - \frac{u_1}{v_1^2}\frac{\mathrm{d}v_1}{\mathrm{d}\rho} \ .$$

After some manipulations, one can write

$$\frac{\mathrm{d}f_{n,i}}{\mathrm{d}\rho} = -f_{n,i}\sum_{j>0} f_{n,j}\Delta_{n,j} + f_{n,i}\Delta_{n,i} \ ,$$

where

$$\Delta_{n,i}(\rho) = \sum_{j=0}^{i-1} \frac{1}{\beta(n,j) + \rho} \ .$$

For the second term we follow the same steps, starting with the definitions

$$u_2(r) = \sum_m m(m-1)s_m g_m \ ,$$

$$v_2(r) = \sum_m m s_m g_m \ ,$$

$$\frac{\mathrm{d}u_2}{\mathrm{d}r} = \sum_m m(m-1)\frac{\mathrm{d}s_m}{\mathrm{d}r}g_m \ ,$$

$$\frac{\mathrm{d}v_2}{\mathrm{d}r} = \sum_m m\frac{\mathrm{d}s_m}{\mathrm{d}r}g_m \ ,$$

such that we can write

$$\frac{\mathrm{d}\rho}{\mathrm{d}r} = \frac{u_2}{v_2} + r\left[\frac{1}{v_2}\frac{\mathrm{d}u_2}{\mathrm{d}r} - \frac{u_2}{v_2^2}\frac{\mathrm{d}v_2}{\mathrm{d}r}\right] \ .$$

We insert the following in the previous definitions

$$\frac{\mathrm{d}s_m}{\mathrm{d}r} = -\frac{m}{(1+mr)^2} \ .$$

Combining all equations, we are able to compute the Jacobian $J_r$ in $\mathcal{O}(m_{\max} + n_{\max}^2)$ elementary operations.

## IV. THRESHOLD MODELS

All quantities are assumed to have reached the stationary sate. For the phase transition analysis we performed in Sec. II B, we assumed that the infection rate $\beta(n, i) > 0 \; \forall i > 0$ to simplify the expressions. Therefore, the results we obtained for the invasion and bistability threshold break down if we use a threshold model, for instance of the form $\beta(n, i) \equiv \lambda \delta_{n-1, i}$, as used in Ref. [6].

To understand why our equations for the critical and tricritical points are not valid, let us consider the two level threshold model

$$\beta(n, i) = H(i - 1)\varepsilon + (\lambda - \varepsilon)\delta_{n-1, i} \; .$$

If $\varepsilon \to 0$, we recover $\beta(n, i) \equiv \lambda \delta_{n-1, i}$. Using Eq. (15), we have for small $\varepsilon$

$$\lambda_c \approx \frac{\langle m \rangle}{\langle m(m - 1) \rangle} \varepsilon^{2-n} \; .$$

Therefore, for $n > 2$ and finite mean excess membership, the invasion threshold diverges in the limit $\varepsilon \to 0$, as can be seen in Supplementary Figure 2. The contagion is then a pure critical mass model [7], i.e., a vanishing initial fraction of infected nodes cannot lead to an endemic state and there is no invasion threshold.

The only way to have a non-diverging invasion threshold when $\varepsilon \to 0$ is to have a diverging excess membership, typically when $g_m \sim m^{-\gamma_m}$. A complete analysis of the critical properties of this regime was performed in Ref. [6] using a heterogeneous mean-field approximation, showing that both continuous and discontinuous phase transitions are possible. The results they obtained are in agreement with our framework. Indeed, if we introduce $\beta(n, i) \equiv \lambda \delta_{n-1, i}$ in Eq. (8), we have

$$f_{n, i} = f_{n, 0} \frac{n!}{(n - i)! i!} \rho^{i-1} (\rho + \lambda \delta_{n, i}) \; ,$$

$$= f_{n, 0} \left\{ \frac{n!}{(n - i)! i!} \rho^i + \lambda \delta_{n, i} \rho^{n-1} \right\} \; ,$$

with

$$f_{n, 0} = [(1 + \rho)^n + \lambda \rho^{n-1}]^{-1} \; .$$

If we further assume that $p_n \equiv \delta_{n, \bar{d}}$ (only one type of interaction), then

$$r = \frac{\lambda \bar{d} \rho^{\bar{d}}}{(1 + \rho)^{\bar{d}} - 1} \approx \lambda \rho^{\bar{d}-1} \; ,$$

for $\rho \ll 1$. This is the same functional form as used in Ref. [6] to describe their mean-field quantities in the stationary state. Our approach would therefore predict the same critical exponents.

## V. UNIFORMLY SAMPLING STUB-LABELED SIMPLE BIPARTITE GRAPHS

The rationale behind the additional $M'$ edge-swap attempts in Sec. IV A 3 is to ensure the uniformity of the generation process. Indeed, we want to sample uniformly from the ensemble of all *stub-labeled* [9] simple bipartite graphs with fixed membership and group size sequences. By performing a stub matching and eliminating the multi-edges, the generating process is not uniform.

However, by using arguments similar to the ones presented in Ref. [9], it straightforward to show that performing random edge swaps as described in the main text creates an ergodic Markov chain that has a uniform stationary distribution on the ensemble of graphs of interest. Thus, if a sufficient number of edge swaps attempts are performed, the Markov chain approximates the stationary distribution and the resulting hypergraph is properly randomized.

To assess whether the chain is sufficiently close to the stationary distribution, it is standard to look at a number of scalar measures and to make sure the chain has converged [10]. In Supplementary Figure 3, we show such results for the bipartite clustering coefficient $C_b$, as defined by Robins and Alexander [8]. We see that for both datasets, whether we start the chain with the original hypergraph or with a random configuration from the stub-matching process, the distribution eventually settles around a stationary value. Moreover, we see that $\sim M'$ edge-swap attempts ($\sim 1$ Markov-chain step per edge) is usually sufficient to reach the stationary distribution, and that using a random configuration from the stub-matching process puts us closer to the stationary distribution initially.

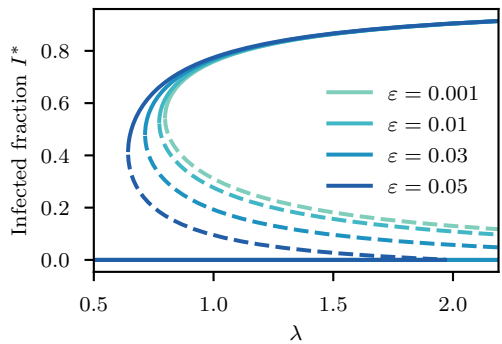

Supplementary Figure 2. **Stationary fraction of infected nodes for a threshold model.** Solid lines and dashed lines represent stable and unstable solutions respectively. The infection rate function is $\beta(n, i) = H(i-1)\varepsilon + (\lambda - \varepsilon)\delta_{n-1,i}$. The structure is fixed with $g_m = \delta_{m,6}$ and $p_n = \delta_{n,3}$.

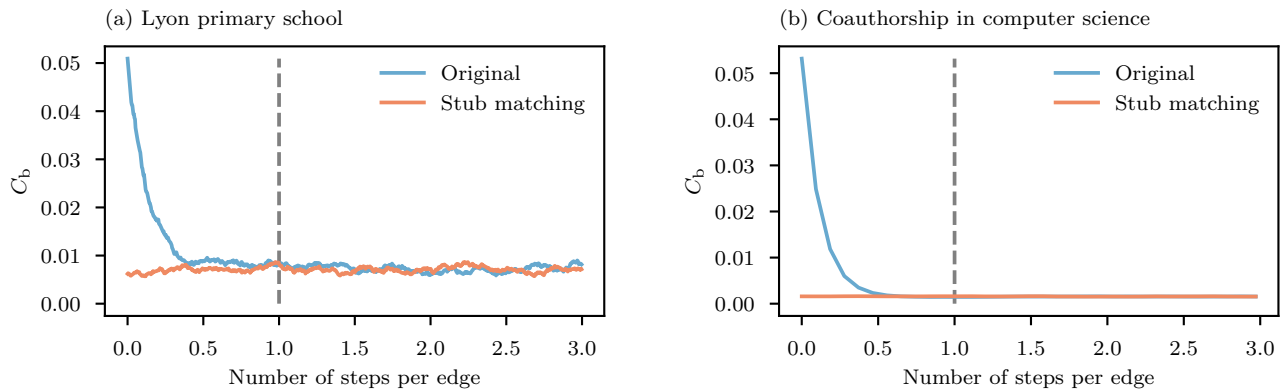

Supplementary Figure 3. **Characterization of the convergence for the randomization procedure.** We define a Markov chain where at each step we perform an edge-swap attempt on the bipartite representation of the hypergraph. If this does not create multi-edges, the step is accepted, otherwise we keep the previous configuration. The stationary distribution of this Markov chain is uniform on all *stub-labeled* simple bipartite graphs. To characterize the mixing time, we look for the convergence of the bipartite clustering coefficient $C_b$, as defined by Robins and Alexander [8]. We initialize the Markov chain with the original hypergraph constructed from the data (blue lines) and a random configuration produced by stub matching (red lines). (a) Hypergraph constructed from high-resolution face-to-face contact data from a primary school in Lyon (see Sec. IV A 2). The bipartite graph representation contains 2853 edges and we perform a measure of the clustering coefficient after each 10 edge-swap attempts. (b) Hypergraph constructed from coauthorship data in computer science (see Sec. IV A 2). The bipartite graph representation contains 215 434 edges and we perform a measure of the clustering coefficient after each $2 \times 10^4$ edge-swap attempts.

## VI. INFLUENCE MAXIMIZATION: LIMIT CASE

All quantities are evaluated at $t = 0$. An important limit case for influence maximization is when the number of nodes to infect $I = \epsilon$ is very small. Fortunately, we can evaluate $\zeta$ in the limit $\epsilon \to 0$. Let us develop $\Phi$ to first order in $\epsilon$ for the optimal solution of both strategies.

For the influential spreader strategy,

$$s_m \simeq 1 - \frac{\epsilon \delta_{m,m_{\max}}}{g_{m_{\max}}},$$

and

$$q \simeq \frac{\epsilon m_{\max}}{\langle m \rangle}.$$

Therefore, to first order in $\epsilon$,

$$\Phi_{\mathcal{S}}^{\star} \propto \frac{\epsilon \langle \beta(n,1) n(n-1) \rangle m_{\max}}{\langle m \rangle} \;.\tag{S3}$$

For the influential group strategy,

$$s_m \simeq 1 - \eta m \;\Rightarrow\; \eta \simeq \frac{\epsilon}{\langle m \rangle} \quad \text{and} \quad u \simeq \epsilon \frac{\langle m(m-1) \rangle}{\langle m \rangle} \;.$$

Also, $\tilde{f}_{n,i} = 0$ for all $n$ and $i > 0$, except for a particular pair $(n', i')$ that maximizes the ratio $R(n,i)$. Equation (28) implies

$$\tilde{f}_{n',i'} \simeq \frac{\epsilon \langle n \rangle}{\langle m \rangle i' p_{n'}} \;,$$

and therefore to first order in $\epsilon$

$$\Phi_{\mathcal{F}}^{\star} \propto \epsilon \frac{\langle n \rangle}{\langle m \rangle} R(n',i') + \epsilon \langle \beta(n,1) n(n-1) \rangle \frac{\langle m(m-1) \rangle}{\langle m \rangle^2} \;,\tag{S4}$$

where more formally

$$R(n',i') = \max\left( \{ R(n,i) | 0 < i \le n \wedge 2 \le n \le n_{\max} \} \right) \;.$$

Finally, combining Eqs. (S3) and (S4) we obtain Eq. (31) in the main text.

---



\end{document}